\documentclass[aps,prx,groupedaddress,notitlepage,showpacs,floatfix,twocolumn]{revtex4-1}
\usepackage{graphicx}
\usepackage{braket}
\usepackage[colorlinks=true,linkcolor=cyan,citecolor=magenta,
filecolor=magenta,urlcolor=cyan,runcolor=cyan]{hyperref}
\usepackage{subfigure}
\usepackage{amsmath,amsfonts,mathrsfs,bbm,bm} 
\usepackage[table,dvipsnames]{xcolor}
\usepackage[capitalize]{cleveref}
\usepackage{lmodern}
\usepackage[T1]{fontenc}
\usepackage{microtype} 
\usepackage{enumitem}

\usepackage{titlesec}

\newcommand*{\ig}[2][width=\columnwidth]{\includegraphics[#1]{#2.pdf}}

\newcommand{\beq}{\begin{equation}}
\newcommand{\eeq}{\end{equation}}
\newcommand{\beqnn}{\begin{equation*}}
\newcommand{\eeqnn}{\end{equation*}}
\newcommand{\bea}{\begin{eqnarray}}
\newcommand{\eea}{\end{eqnarray}}
\newcommand{\beann}{\begin{eqnarray*}}
\newcommand{\eeann}{\end{eqnarray*}}
\newcommand{\bes}{\begin{subequations}}
\newcommand{\ees} {\end{subequations}}
\newcommand*{\ensuretext}[1]{\ifmmode \mathrm{#1} \else #1 \fi}
\newcommand*{\pudenz}{\ensuremath{{[3,1,3]}_{1}}\ }
\newcommand*{\pudenzcode}{\ensuremath{{[3,1,3]}_{1}} \ensuretext{code}}
\DeclareMathOperator*{\argmax}{arg\,max}

\newcommand{\mc}[1]{\mathcal{#1}}

\makeatletter
\@addtoreset{section}{part}
\makeatother
\titleformat{\part}[display]
{\normalfont\LARGE\bfseries\centering}{}{0pt}{}

\begin{document}

\title{Analog Errors in Quantum Annealing: Doom and Hope}

\author{Adam Pearson$^{(1,2)}$, Anurag Mishra$^{(1,2,6)}$,
Itay Hen$^{(1,2,3)}$, Daniel A. Lidar$^{(1,2,4,5)}$}
\affiliation{
$^{(1)}$Department of Physics and Astronomy, 
$^{(2)}$Center for Quantum Information Science \& Technology, 
$^{(3)}$Information Sciences Institute, 
$^{(4)}$Department of Electrical Engineering, 
$^{(5)}$Department of Chemistry, University of Southern California, Los Angeles, CA 90089\\
$^{(6)}$Current address: Qulab Inc., 1642 Westwood Blvd, Los Angeles, CA 90024, USA}

\begin{abstract}
Quantum annealing has the potential to provide a speedup over classical algorithms in solving optimization problems. Just as for any other quantum device, suppressing Hamiltonian control errors will be necessary before quantum annealers can achieve speedups. Such analog control errors are known to lead to $J$-chaos, 
wherein the probability of obtaining the optimal solution, encoded as the ground state of the intended Hamiltonian, varies widely depending on the control error.
Here, we show that $J$-chaos causes a catastrophic failure of quantum annealing, in that the scaling of the time-to-solution metric becomes worse than that of a deterministic (exhaustive) classical solver. We demonstrate this empirically using random Ising spin glass problems run on the two latest generations of the D-Wave quantum annealers. We then proceed to show that this doomsday scenario can be mitigated using a simple error suppression and correction scheme known as quantum annealing correction (QAC). By using QAC, the time-to-solution scaling of the same D-Wave devices is improved to below that of the classical upper bound, thus restoring hope in the speedup prospects of quantum annealing.
\end{abstract}

\maketitle

\section{Introduction}
\label{sec:introduction}

\setlength{\parskip}{2mm plus0.5mm minus0.5mm} 

The demonstration of scaling speedups~\cite{Shor:97,Bravyi:2017aa} using quantum hardware is the holy grail of quantum computing, and massive efforts are underway worldwide in their pursuit. A daunting obstacle is the fact that all physical implementations of quantum computers suffer from analog control errors, in which the coefficients of the Hamiltonian implemented differ from those intended (e.g.,~\cite{barends_2014_superconductingquantum}), a fact that threatens to spoil the results of computations due to the accumulation of small errors. This problem was recognized early on in the gate model of quantum computing~\cite{Landauer:95}, and soon after theoretically dealt with by error discretization via quantum error correcting codes~\cite{Shor:96}. Moreover, the accuracy threshold theorem guarantees that if the physical gates used to implement encoded, error-corrected quantum circuits have sufficiently high fidelity, then any real, noisy quantum computation can be made arbitrarily close to the intended, noiseless computation with modest overhead~\cite{Aliferis:05,Chao:2017aa,Lidar-Brun:book}. In contrast, the ultimate impact of analog control errors in Hamiltonian quantum computing, in particular in adiabatic quantum optimization~\cite{Farhi:00} and quantum annealing (QA)~\cite{kadowaki_quantum_1998}, is not as clear. While unlike the gate model adiabatic quantum evolution is inherently robust to path variations due to unitary control errors~\cite{childs_robustness_2001}, there is as of yet no equivalent mechanism of error discretization or an analogous accuracy threshold theorem in this paradigm. Yet, at the same time quantum annealing offers a currently unparalleled opportunity to explore NISQ-era~\cite{Preskill:2018aa} quantum optimization with thousands of qubits~\cite{Albash:2017aa,Mandra:2017ab}. It is thus of great importance to assess the role of analog control errors in QA, and to find ways to mitigate them. Here we do so in the context of spin glass problems, which are known to exhibit a type of control-error induced bond or disorder chaos already in a purely classical setting, causing chaotic changes in the ground or equilibrium state~\cite{Bray:1987,katzgraber_2007_disorder}.

The extent to which analog control errors present a challenge in using any physical realization of quantum annealing, such as the D-Wave processors~\cite{Bunyk:2014hb,harris_flux_qubit_2010,Harris:2010kx}, for optimization, cannot be overstated. Indeed, an earlier study of such errors in these processors found evidence of sub-classical performance and referred to the effect as $J$-chaos~\cite{Martin-Mayor:2015dq}, a terminology we adopt here. More recently it was shown that analog control noise causes a decrease in the probability that the implemented Hamiltonian shares a ground state with the intended Hamiltonian that scales exponentially in the size of the problem and the magnitude of the noise~\cite{albash_2019_analogerrors}. This means that even if the annealer solves the implemented problem correctly, it has an exponentially shrinking probability of finding the intended ground state. In other words, subject to $J$-chaos an otherwise perfectly functioning quantum annealer will typically find the correct answer to the wrong problem.

To mitigate this ``wrong Hamiltonian'' problem and restore the prospects for a speedup in the use of quantum annealing for optimization, it is necessary to introduce techniques for error suppression and correction. This observation is not new~\cite{Young:2013fk,Bian:2014,Zhu:2015pd}, and repetition coding along with the use of energy penalties has been shown to significantly enhance the performance of quantum annealers~\cite{PAL:13,PAL:14,Mishra:2015,Vinci:2015jt,vinci2015nested,Vinci:2017ab}. In contrast to previous work, here, for the first time, we directly address the impact of $J$-chaos on algorithmic scaling of optimization in experimental quantum annealing, while accessing a computational scale that is still far out of reach of current gate-model quantum computing devices. We employ a quantum annealing correction method to mitigate the problem, and demonstrate that while the scaling of uncorrected quantum annealing in solving random Ising spin glass problems is catastrophically affected by $J$-chaos --- in that it is worse than even that of a deterministic (brute force) classical solver --- hope for a quantum speedup~\cite{speedup} is restored with error suppression and correction. This reassuring conclusion is reached here using the simplest possible error suppression and correction scheme~\cite{PAL:13}, so that much room for improvement remains for more advanced methods. We expect our results to apply broadly, certainly beyond the D-Wave devices to other quantum~\cite{Weber:2017aa,Novikov:2018aa,Goto:2019aa} and semiclassical annealing implementations~\cite{Inagaki:2016aa,Goto:2019ab}, and to other forms of analog quantum computing~\cite{RevModPhys.80.1061}.

\section{Results}
\label{sec:results}

Inspired by classical simulated annealing, in which thermal fluctuations are used to hop over barriers,  quantum annealing uses quantum fluctuations to tunnel through  barriers~\cite{kadowaki_quantum_1998,PhysRevB.39.11828,Brooke1999,Santoro,Boixo:2014yu,Muthukrishnan:2015ff,PhysRevX.6.031015}. The D-Wave processors are physical implementations of such devices~\cite{Dwave}. In the standard forward annealing protocol they apply a time-dependent transverse field $H_X = \sum_{i} \sigma^{x}_{i}$ ($\sigma^{a}_{i}$ denotes the Pauli matrix of type $a\in\{x,y,z\}$ applied to qubit $i\in\{1,\dots,N\}$) and Ising Hamiltonian as follows,
\begin{equation}
  \label{eq:1}
  H(s) = A(s) H_X + B(s) \tilde{H}_{\mathrm{Ising}} ,
\end{equation}
where $\tilde{H}_{\mathrm{Ising}} = H_{\mathrm{Ising}} + \delta H_{\mathrm{Ising}}$ and $H_{\mathrm{Ising}}$ are, respectively, the {implemented} (perturbed) and {intended} (unperturbed) ``problem'' Hamiltonians, while $\delta H_{\mathrm{Ising}}$ is an error term (the perturbation). 
Thus $\tilde{H}_{\mathrm{Ising}}$ is the (wrong) Hamiltonian including analog control errors while $H_{\mathrm{Ising}}$ is the Hamiltonian whose ground state we wish to find as the solution to the optimization problem specified by a set of local fields $\{h_i\}$ and couplers $\{J_{ij}\}$:
\bes
\begin{align}
  \label{eq:2}
  H_{\mathrm{Ising}} &= \sum_{i\in\mathcal{V}}h_{i}\sigma^{z}_{i} + \sum_{(i,j)\in\mathcal{E}} J_{ij} \sigma^{z}_{i}\sigma^{z}_{j}\\
  \delta H_{\mathrm{Ising}} &= \sum_{i\in\mathcal{V}}\delta h_{i}\sigma^{z}_{i} + \sum_{(i,j)\in\mathcal{E}} \delta J_{ij} \sigma^{z}_{i}\sigma^{z}_{j} ,
\end{align}
\ees
where $\mathcal{V}$ and $\mathcal{E}$ are the vertex and edge sets of the graph $\mathcal{G}$, and $N = |\mathcal{G}|$.
We assume that the noise is Gaussian with zero mean and standard deviation $\eta$:
\beq
\delta h_i,\delta J_{ij} \sim \mc{N}(0,\eta^2) .
\label{eq:deltas}
\eeq
We note that analog errors resulting in the replacement of $H_X$ by $\sum_i (1+\epsilon_i)\sigma^x_i$ with random $\epsilon_i$ are expected as well, but we do not consider such transverse field errors here.
The normalized time, $s=t/t_\mathrm{f}$, with $t_\mathrm{f}$ denoting the final time, increases from $0$ to $1$, with $A(0) \gg B(0)$ and $B(1) \gg A(1)$, and $A(s)$ [$B(s)$] decrease (increase) monotonically. As such, the transverse field initially drives strong quantum fluctuations that eventually give way to the implemented Ising Hamiltonian. In the absence of $\delta H_{\mathrm{Ising}}$ and an environment the adiabatic theorem guarantees that the ground state of $H_{\mathrm{Ising}}$ will be found if $t_\mathrm{f}$ is large compared to the inverse of the minimum gap of $H(s)$ and the maximum time-derivative(s) of $H(s)$~\cite{Jansen:07,lidar:102106}. In the presence of an environment the adiabatic theorem instead guarantees evolution towards the steady state of the corresponding Liouvillian~\cite{Avron:2012tv,Venuti:2015kq}, which becomes the ground state only for sufficiently low temperature (compared to the gap of the Liouvillian). When $\delta H_{\mathrm{Ising}} \neq 0$, the probability that the computation ends in the ground state of $H_{\mathrm{Ising}}$ decreases exponentially in both $N$ and some power of $\eta$~\cite{albash_2019_analogerrors}. The reason is that as more noise is added, there is an increasing probability that the spectrum changes such that the ground state is swapped with an excited state. Thus, increasing noise and problem size leads to a rapidly growing probability of failure to find the correct ground state. In experimental quantum annealing both the environment and control errors inevitably play a role.

\begin{figure}[b]
  \centering
  \includegraphics[width=\columnwidth]
    {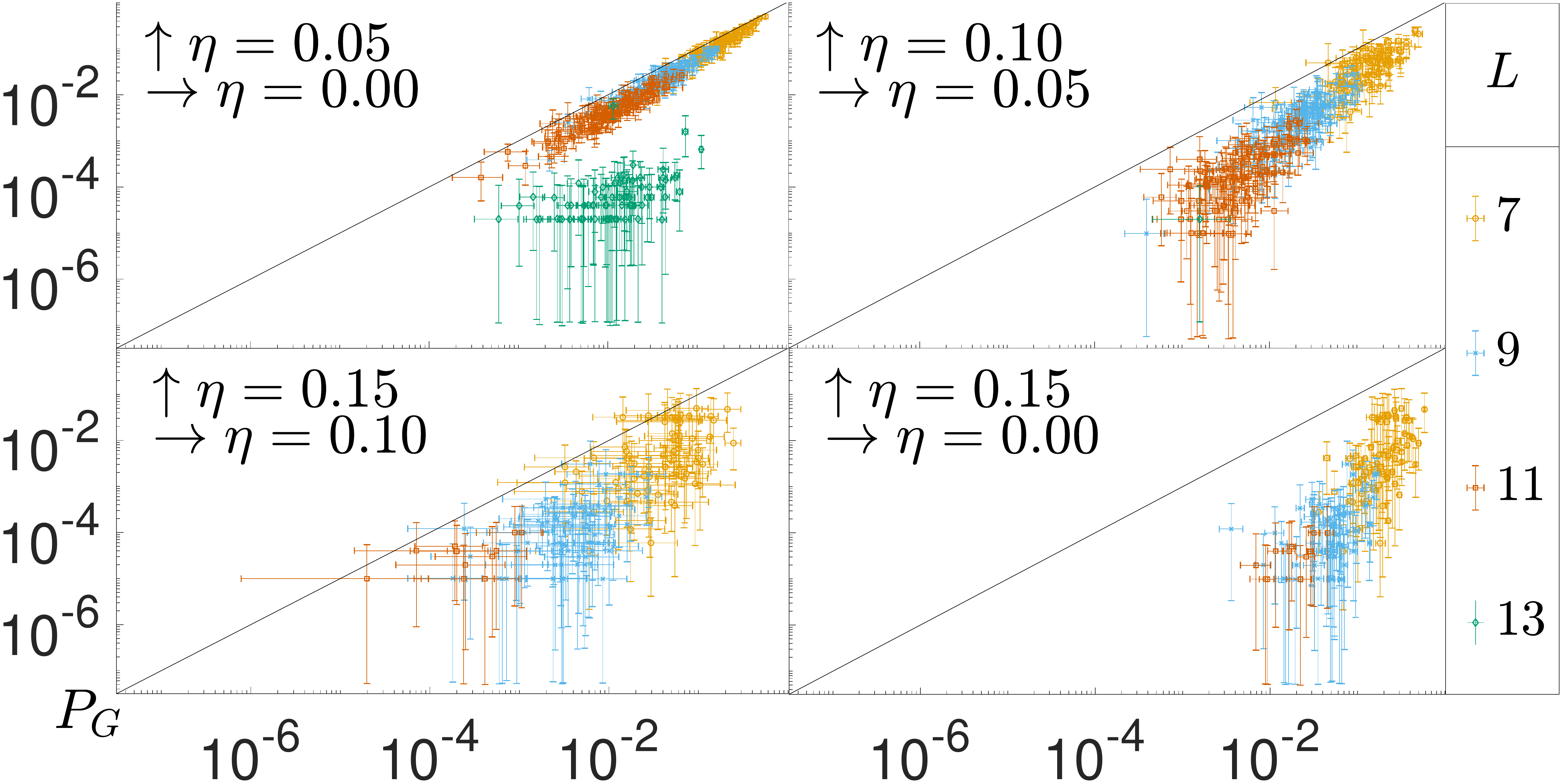}
  \caption{\textbf{Correlation plots of ground state (success) probability $P_g$ with increasing added
    noise $\eta$.} Results for random Ising instances are shown after addition of coupler noise $\delta J_{ij} \sim \mc{N}(0,\eta^2)$. In each panel every data point is $P_g$ for the same instance with different $\eta$ values.  Top left: success probability
    decreases as we go from the no added noise case $\eta=0$ to $\eta=0.05$. Note that even for $\eta=0$ there is an intrinsic control error $\eta_{\mathrm{int}}$. The same
    trend persists as we compare $\eta=0.05$ to $\eta=0.10$ (top right) and $\eta=0.10$ to $\eta=0.15$ (bottom left). The bottom right panel shows the extreme
    comparison of $\eta=0$ to $\eta=0.15$: the ground state is almost never found for $\eta=0.15$. In all plots success probability generally decreases with increasing size $L$. Instances with zero success
    probability are not shown, hence the number of data points drops with increasing $\eta$. Unless otherwise noted, here and in later plots error bars always denote 95\% confidence intervals (C.I.) obtained via a bootstrap. Details of our data analysis are given in Methods~\cref{sec:data-analysis}.}
  \label{fig:successes}
\end{figure}

\subsection{Effect of control noise}
\label{sec:ecn}

In order to systematically test the effect of analog control errors we studied the performance of two D-Wave devices on random Ising instances of varying size $N$, to which we added artificially generated Gaussian control noise $\eta$. This noise was added to the intrinsic analog device noise $\eta_{\mathrm{int}}$ \footnote{The D-Wave documentation refers to this as Integrated Control Errors (ICE). According to~\url{https://docs.dwavesys.com/docs/latest/c_qpu_1.html}, $\eta_{\mathrm{int}}\lesssim 0.015$ for the DW2000Q device.}, so that the total control noise had variance $\eta^2_{\mathrm{int}}+\eta^2$. Adding noise in this manner allowed us to test its effect on algorithmic performance, and the efficacy of the quantum annealing correction (QAC) strategy described below.

Throughout this work we used Ising instances with local fields $h_{i} = 0$ and couplers $J_{ij}$ selected uniformly at random from the set $\pm \{1/6, 1/3, 1/2\}$, and chose $\eta\in\{0,0.03,0.05,0.07,0.10,0.15\}$. Such instances have been studied before~\cite{q108,speedup,PAL:14,2014Katzgraber,King:2014uq}, but only with $\eta=0$, i.e., never subject to the systematic addition of control noise. We define ``success'' in all cases as finding a ground state of the unperturbed Hamiltonian $H_{\mathrm{Ising}}$. See Methods~\cref{sec:insts} for a complete description of the instances and how we verified ground states.

The number of qubits $N$ is proportional to $L^2$, the number of Chimera graph unit cells of the D-Wave devices we used (see Methods~\cref{sec:d-wave}).
Figure~\ref{fig:successes} displays a series of correlation plots between different levels of added noise, for different problem sizes parametrized by $L$.
At every size, addition of noise results in a lower success probability for all instances. Increased size also results in lower success probability, as expected. Thus, Fig.~\ref{fig:successes} gives a visual confirmation of the detrimental effect of control noise; we quantify this systematically below.

\begin{figure}[b]
  \centering
  \includegraphics[width=\columnwidth]
    {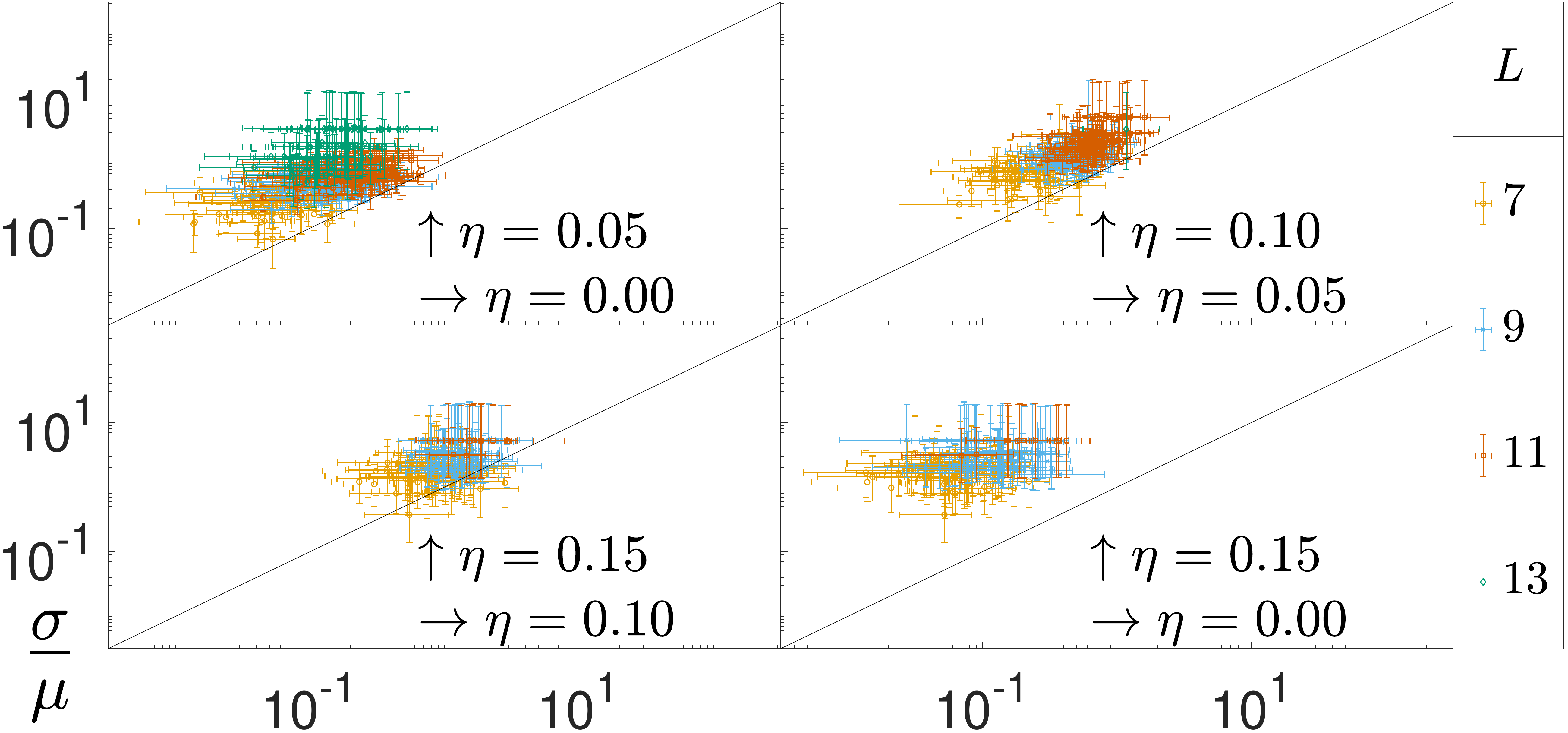}
  \caption{\textbf{Correlation plots of J-chaoticity measure $\sigma/\mu$  with increasing added
    noise $\eta$.}  Details are as in~\cref{fig:successes}, but here we compare the J-chaoticity measure of instances with different added noise values. The measure
   increases as we go from the no added noise case $\eta=0$ to $\eta=0.05$ (top left). The same trend persists as we compare $\eta=0.05$ to $\eta=0.10$ (top right) and $\eta=0.10$ to $\eta=0.15$ (bottom left). The bottom right panel shows the extreme comparison of $\eta=0$ to $\eta=0.15$, where for all instances at all sizes J-chaoticity is higher to well within the 95\% C.I.s}
  \label{fig:perfs}
\end{figure}

\begin{figure}[t]
  \centering
  \includegraphics[width=\columnwidth]
    {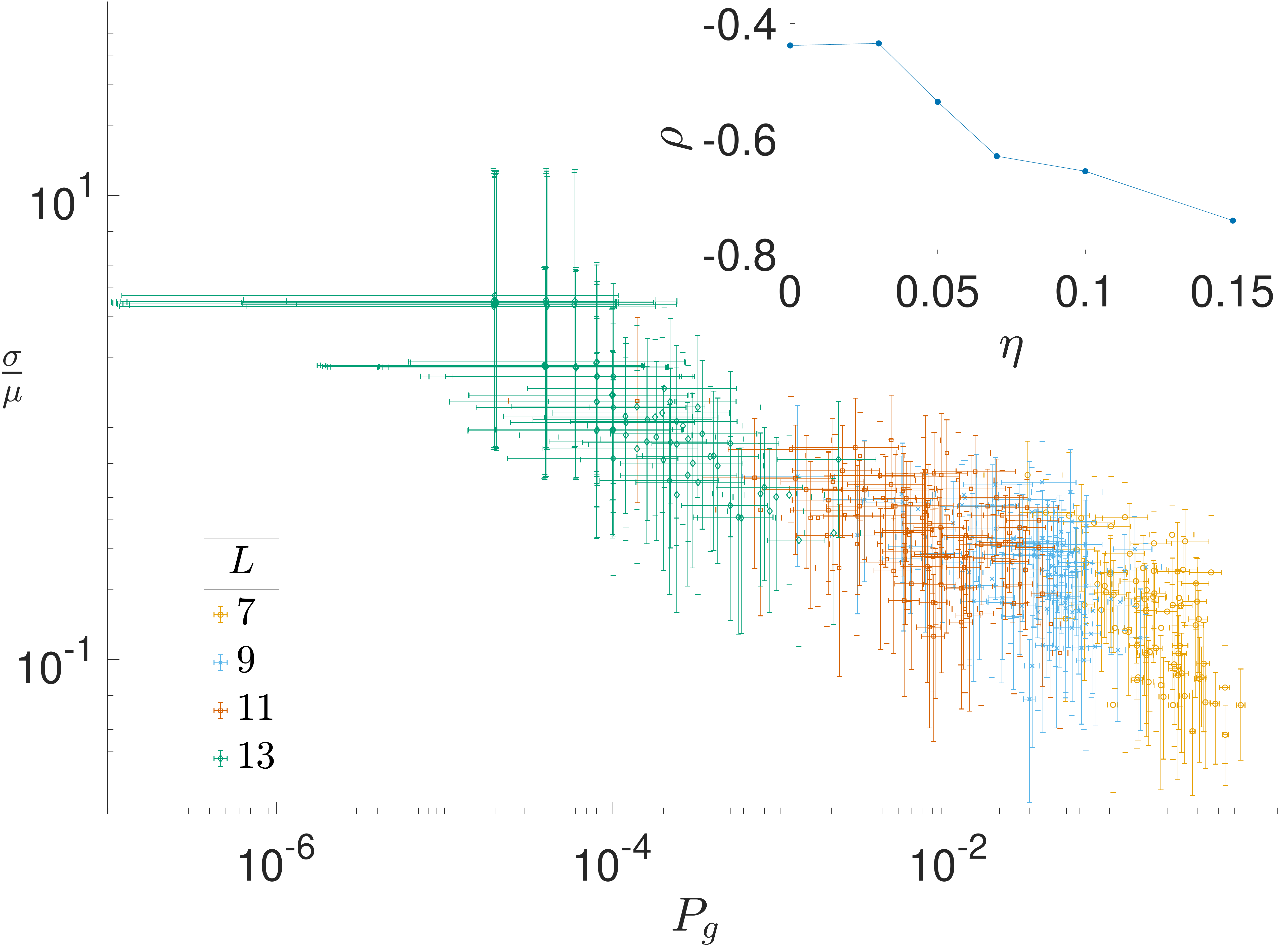}
  \caption{\textbf{Correlation between success probability and J-chaoticity.}
  We plot the success probability and J-chaoticity measure with
    noise $\eta=0.03$. There is evidently a strong correlation between the two, showing  that the success probability is lowered due to $J$-chaos. The inset shows the Pearson correlation coefficient $\rho$ between success probability and J-chaoticity, as a function of added noise, for all $\eta$ values tested. As $\eta$ grows, success probability and J-chaoticity become more negatively correlated.}
  \label{fig:success_perf}
\end{figure}

Next we test whether control noise results in $J$-chaos. The latter exhibits itself as large variations in the success
probability of the programmed Hamiltonian across different runs. We quantify this in terms of the J-chaoticity measure
$\sigma/\mu$, where $\sigma$ is the standard deviation of the success probability
across repeated runs of a given instance, and $\mu$ is the corresponding mean (see Methods~\cref{sec:data-analysis} for more details). In~\cref{fig:perfs}, we plot the correlation of the J-chaoticity measure with increasing noise. For most instances this quantity becomes larger with increasing noise and size, which indicates that
they are becoming more chaotic. Success probability is also strongly (negatively) correlated with increasing J-chaoticity, as shown in~\cref{fig:success_perf}. This establishes that control-noise induced $J$-chaos is responsible for a strong decline in performance. Before we quantify this decline in terms of the time-to-solution metric, we first address how to mitigate this problem using error suppression and correction.

\subsection{Quantum Annealing Correction}
\label{sec:qac}

The error suppression and correction scheme used in this work is the \pudenz QAC code
introduced in~\cite{PAL:13} and further studied experimentally in~\cite{PAL:14,Vinci:2015jt,Mishra:2015}. We refer the reader to these references for details, and to Methods~\cref{sec:QAC} for a brief summary. The \pudenz code is a three-qubit repetition code that corrects bit-flip errors, with the subscript denoting one extra penalty qubit. The penalty term energetically suppresses all errors that do not commute with $\sigma^z$ during the anneal.

Since the \pudenzcode graph is a minor of the Chimera graph (see Methods~\cref{fig:logical-graphs}), we can also implement these instances without QAC. But, to ensure a fair comparison we need to equalize the resources used with QAC and without it. The \pudenzcode consumes four qubits to encode one logical qubit. Thus, we can use the same amount of resources as the encoded logical problem by running four unencoded copies in parallel, which is called the classical repetition strategy (C). To be clear, the difference between QAC and the C strategy is fourfold: QAC uses logical qubits, logical operators, and an energy penalty term, while C uses physical qubits, physical operators, and no penalty. The decoding strategy for QAC is a majority vote over the three data qubits of each logical qubit, while for C it is best-of-four-copies of the logical problem solved by QAC. A more powerful, nested QAC strategy is known~\cite{vinci2015nested}, but it requires more physical qubits per logical qubit, and hence is less suitable for a scaling analysis of the type we perform here.

\begin{figure}
  \centering
  \includegraphics[width=\columnwidth]{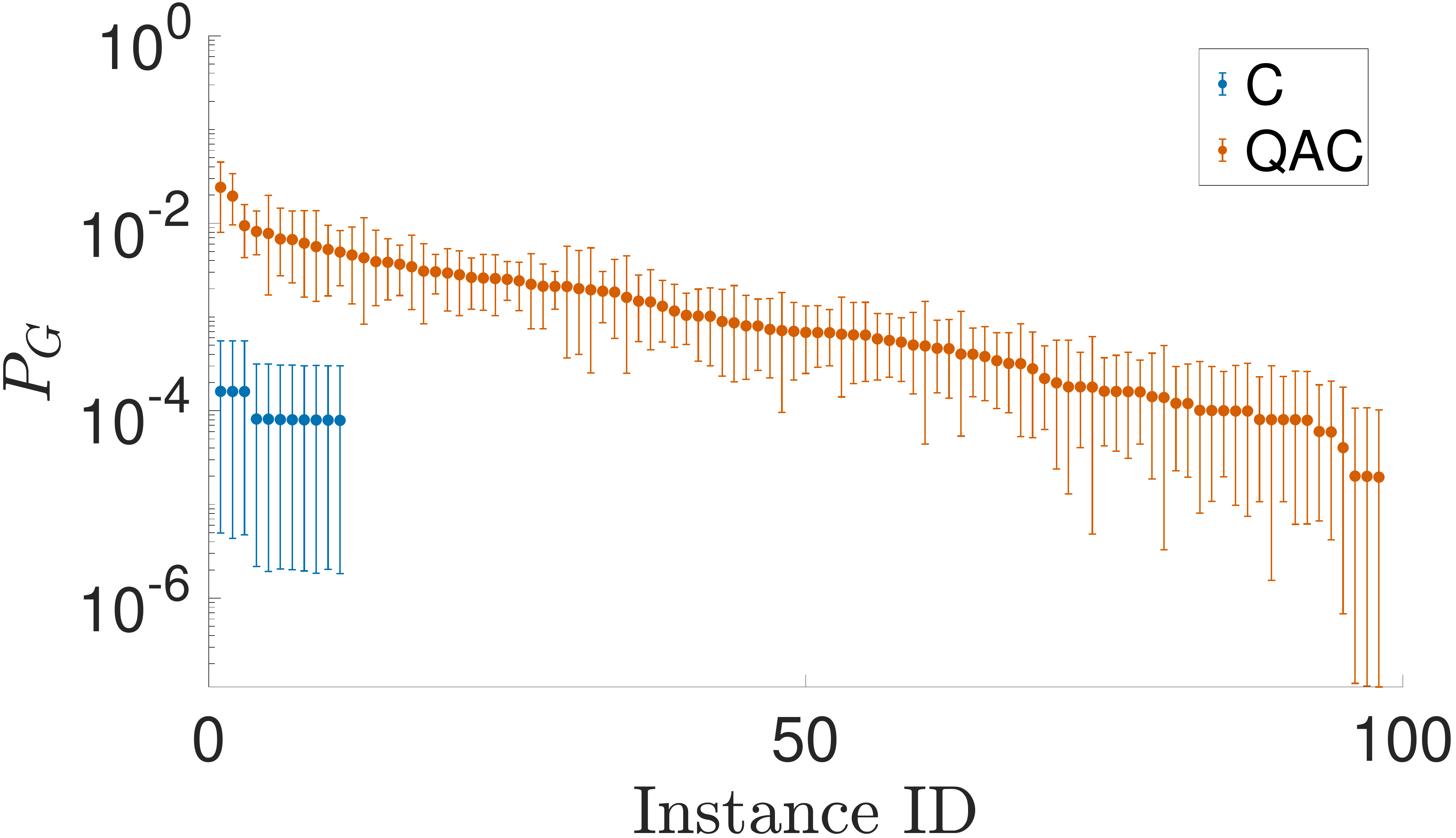}
  \caption{\textbf{Sorted success probabilities for $L=14$ and $\eta=0.07$, all $100$ instances}. C is able to find the correct ground state for only $11$ instances. Meanwhile, QAC is able to find the correct ground state for all but $2$ instances and increases the mean success probability over C.}
  \label{fig:success_14_0.07}
\end{figure}

We now discuss the results after the application of QAC and compare them to the C strategy.
\cref{fig:success_14_0.07} illustrates that for relatively large problem sizes and strong added control noise, such as at $L = 14$ and $\eta = 0.07$, QAC is able to find nearly all ground states while the C strategy only finds the ground state of a small fraction of instances.

\begin{figure}
  \centering
  \subfigure[]{\includegraphics[width=\columnwidth]
   {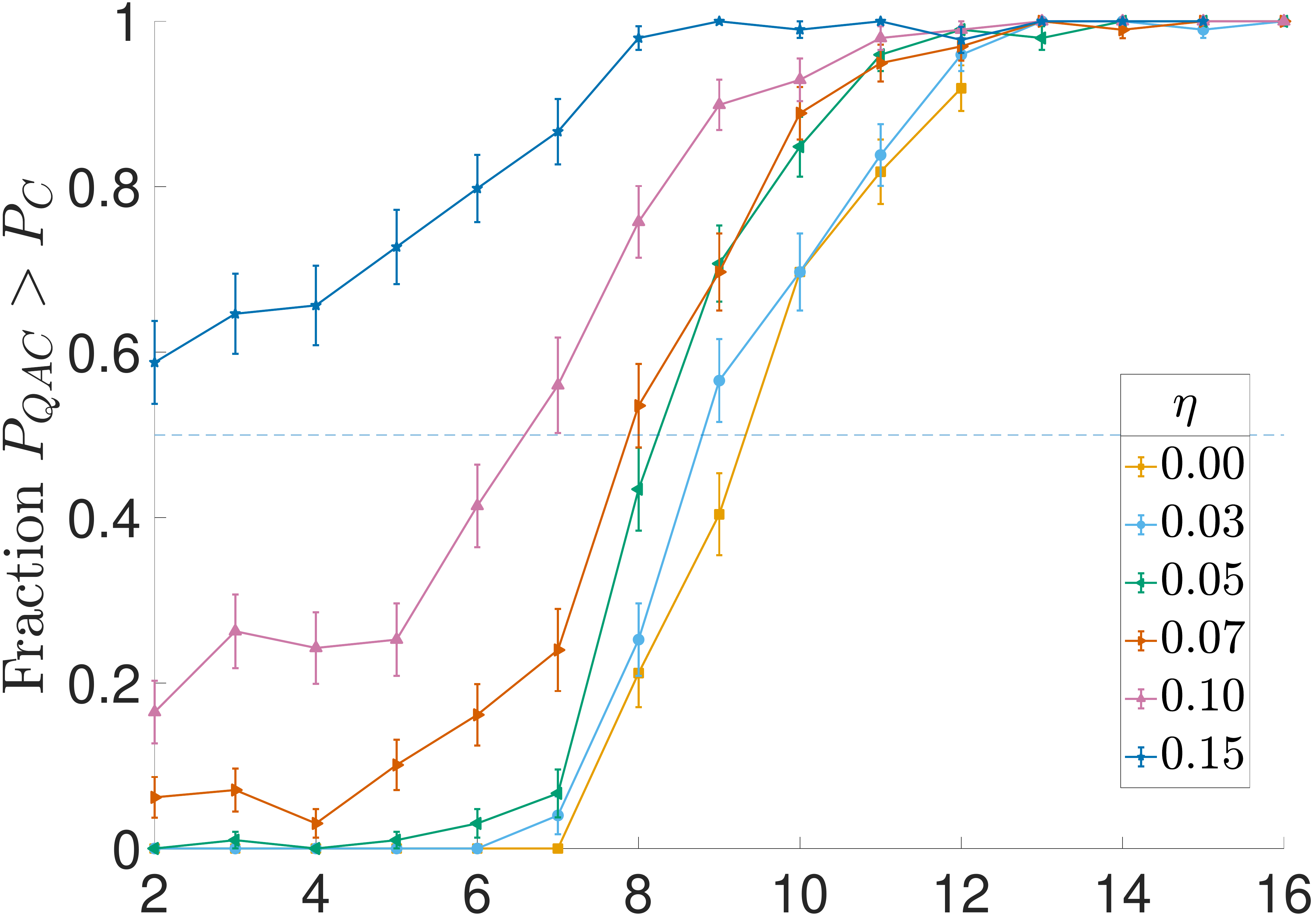}\label{fig:betters_success}}
  \subfigure[]{\includegraphics[width=\columnwidth]
   {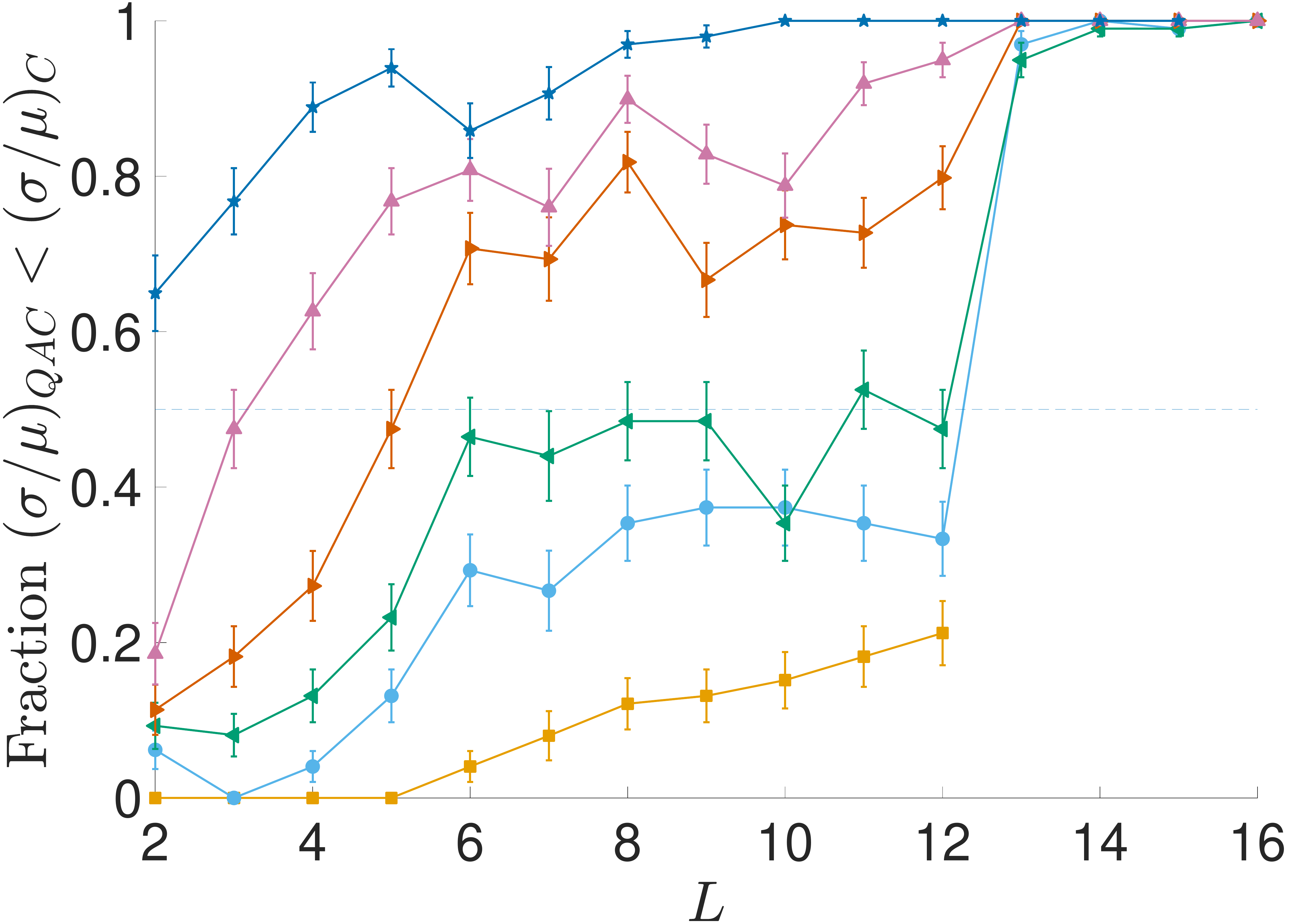}\label{fig:betters_perf}}
  \caption{\textbf{Fraction of instances where QAC outperforms the C
      strategy.} In (a), we plot the fraction of instances where QAC found
    higher success probability than the C strategy after removing any instances where both failed completely (see~\cref{app:fail}).
     In (b), we plot the fraction of instances where QAC lowered the J-chaoticity measure when compared to the C strategy. QAC becomes better with increasing size. At large noise, QAC becomes the better strategy for all instances. We note that here and the other figures below, we have omitted the $\eta =0$ results for $L\geq 13$. The reason is a discontinuity between the DW2X and DW2000Q devices which is discussed in detail in~\cref{app:difference}, and which is unrelated to the scaling analysis that forms the main focus of this work. }
  \label{fig:betters}
\end{figure}

More systematically, we show in~\cref{fig:betters_success} the fraction of instances where using QAC improved success
probability when compared to the C strategy. If more than half of the instances exhibit
better performance for the QAC strategy, applying it is useful for median
instances. Evidently, QAC becomes a better strategy for large size and large noise, i.e., as finding the ground state becomes harder. Similarly, we compare the J-chaoticity measure
$\sigma/\mu$ for QAC and C, as seen in ~\cref{fig:betters_perf}.
Just as in~\cref{fig:betters_success}, we see greater advantage using QAC over C with increasing size and noise.
These observations are consistent with earlier
results~\cite{PAL:13,PAL:14,Vinci:2015jt,Mishra:2015} where the \pudenzcode 
performs better than the classical repetition strategy at large problem sizes. Here, we
have shown that this is also true in the high control noise and $J$-chaos regime.

To understand the source of the improvement in the success probability, consider that application of
repetition codes can decrease the effective noise on the encoded
problem~\cite{Young:2013fk}. In particular, the encoded operators of an $n$-qubit repetition code
have an effective energy scale that scales extensively relative to the unencoded problem (see Methods~\cref{sec:QAC}), while the
random control noise adds up incoherently and hence its energy contribution only scales up by a factor of $\sqrt{n}$. Thus, the encoding reduces the
effective noise of the problem by the factor $1/\sqrt{n}$. This enhances the success probability of
the encoded problem over the simple C strategy, essentially by leveraging just classical properties of the repetition code. However, there is a quantum mechanism at work as well: a mean-field analysis reveals that the penalty term reduces the tunneling barrier width and height in the QAC case~\cite{MNAL:15,Matsuura:2016aa}. Indeed, we shall next see that our empirical results are inconsistent with the constant success probability enhancement that would be expected from a purely classical reduction of the effective noise by the factor $1/\sqrt{3}$ (we use an $n=3$ repetition code).

\subsection{Scaling of the time-to-solution}
\label{sec:scaling-rts}

\begin{figure}
  \centering
  \subfigure[\ C]{\ig[width=\columnwidth]
    {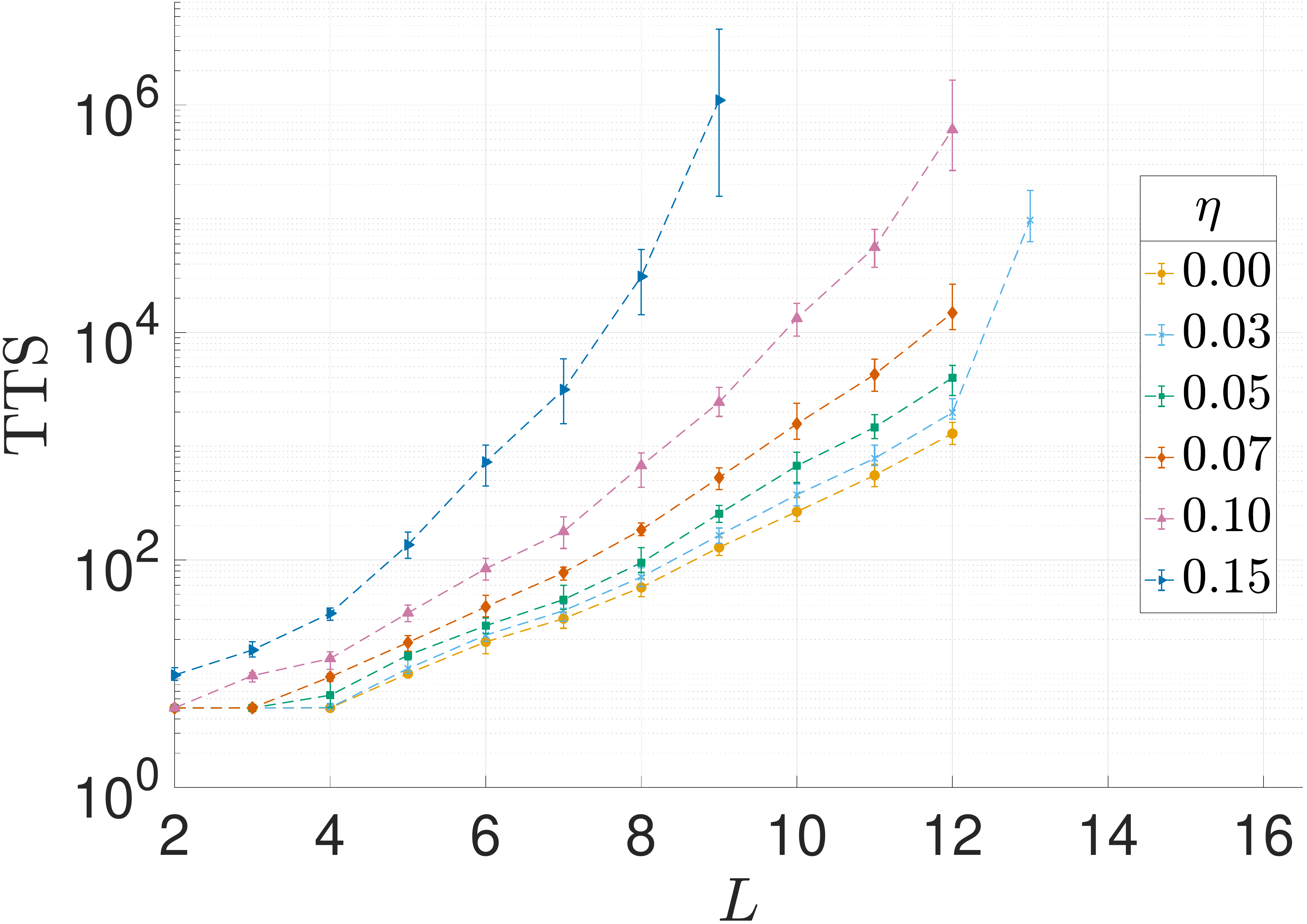}\label{fig:rts_noise_C}}
  \subfigure[\ QAC]{\ig[width=\columnwidth]
    {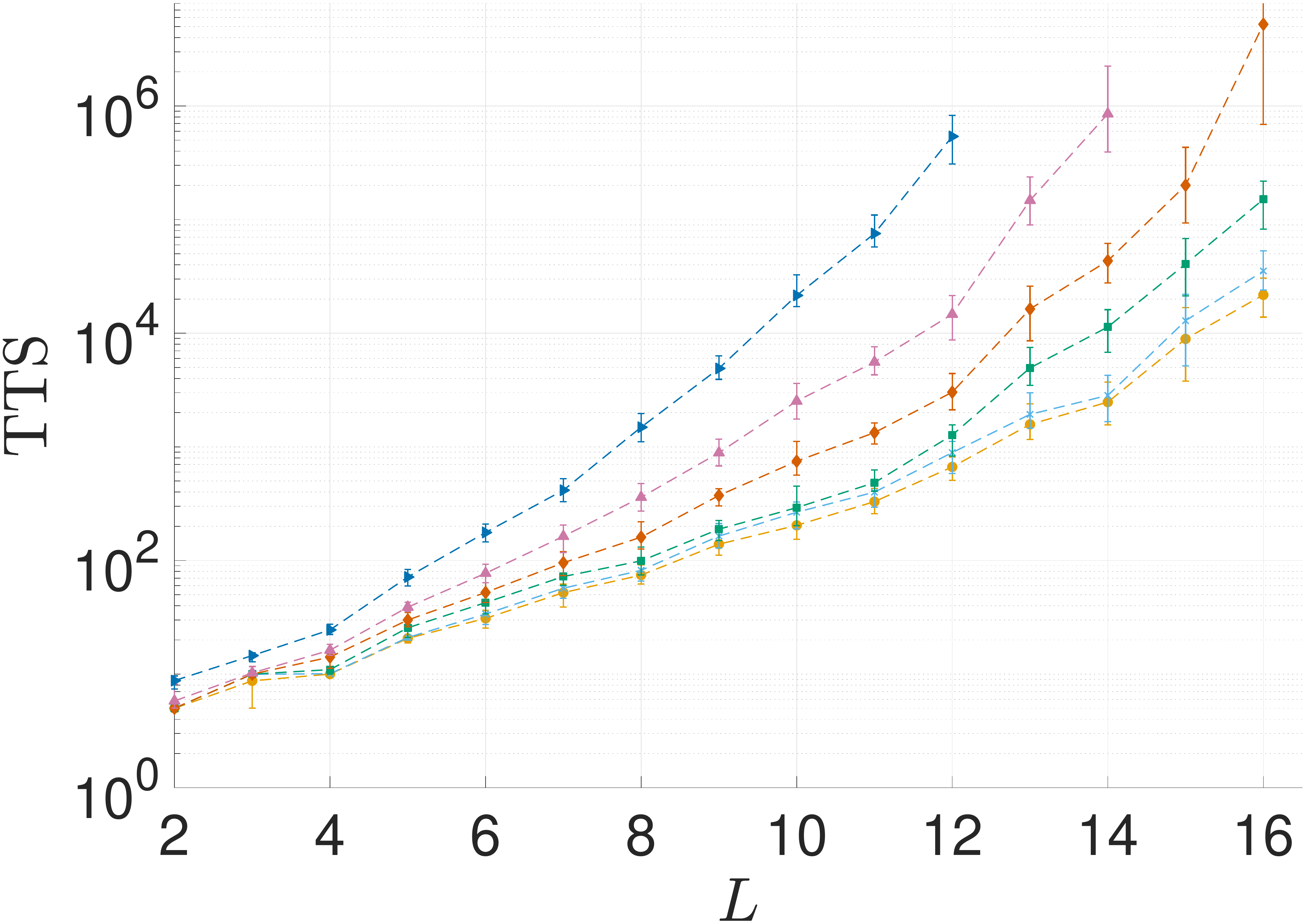}\label{fig:rts_noise_QAC}}
  \caption{\textbf{TTS for QAC and the C strategy, sorted by added noise.} We show the TTS to find at least one ground state of the median instance
    of class $(L,\eta)$. In (a), we show the results from the classical repetition
    strategy. For large instances $L\geq10$ and high noise $\eta=0.15$, the C
    strategy fails to find any ground state for the median instance in our data set. In
    (b), we show the results for QAC, which is always able to find
    the ground state of the median instance in our data set. The scaling in (b) is milder
    compared to (a). For all $\eta>0$, a missing data point indicates that the ground state was never found at that $L$ value, e.g., for all $L\geq 14$ in (a).
}
    \label{fig:rts_noise}
\end{figure}

\begin{figure}
  \centering
  {\includegraphics[width=\columnwidth]
    {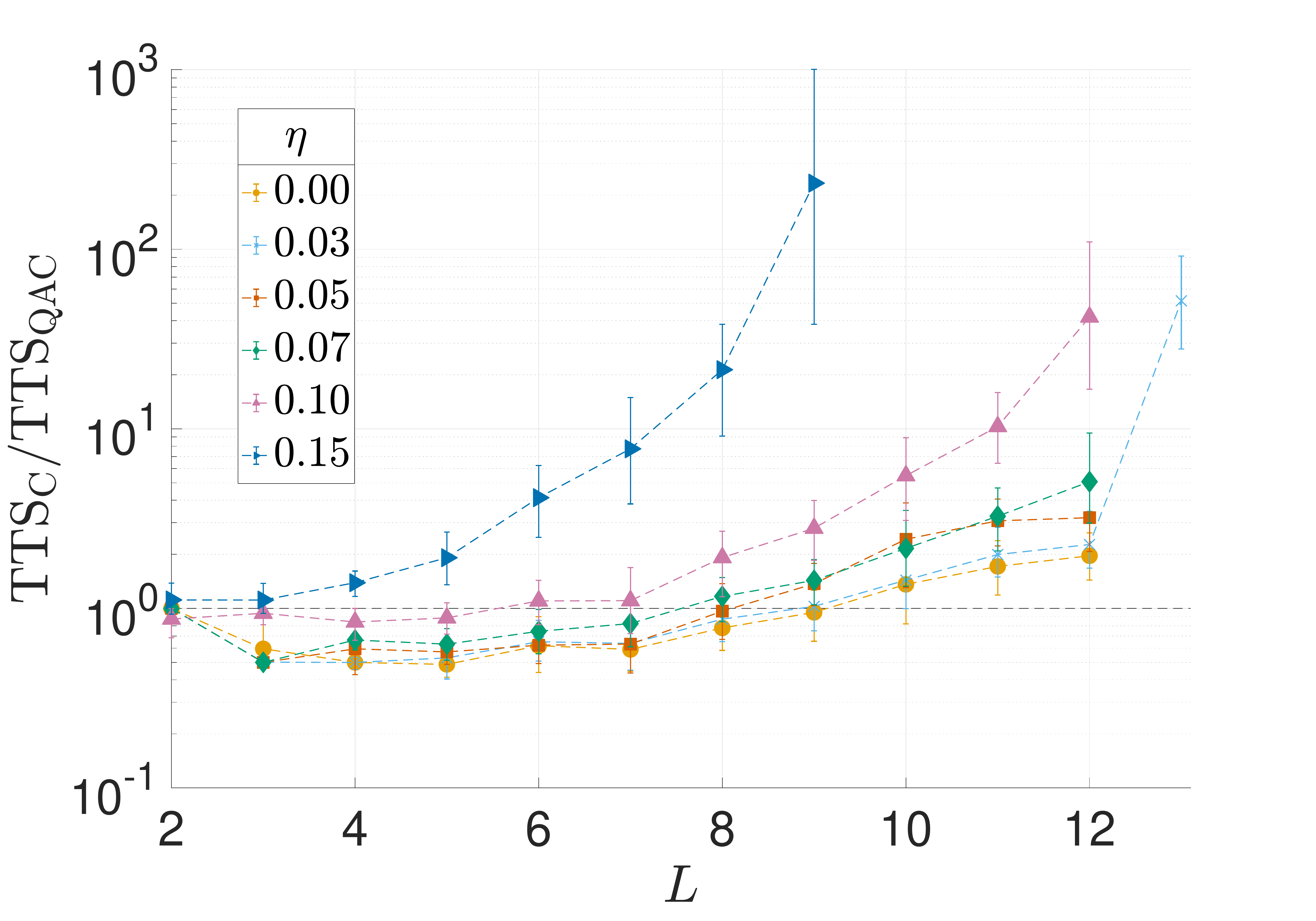} }
  \caption{\textbf{Median speedup of QAC over C.} We show the ratio of TTS of C and TTS of QAC, which will sit above the dashed line at speedup ratio = 1 for cases when it is advantageous to use QAC. Furthermore, a positive slope indicates a potential scaling speedup of QAC over C, since this occurs when the TTS of C is growing more quickly than of QAC. }
   \label{fig:speedup}
\end{figure}

We now discuss the impact of the analog noise on the computational effort required to find
a solution of the problem. This can be quantified by the time-to-solution (TTS) metric~\cite{speedup},
which is the number of runs required to obtain the correct ground state at least once with 99\% success probability:
\begin{equation}
  \label{eq:8}
  \mathrm{TTS} = t_\mathrm{f}\left\lceil \frac{\ln(1-0.99)}{\ln(1-P_g)} \right\rceil \ ,
\end{equation}
where $t_\mathrm{f}$ is the total anneal time per run, and $P_g$ is the probability of finding the ground state in a given run (see Methods~\cref{sec:data-analysis} for details on how $P_g$ was computed). The TTS metric is often used for
benchmarking quantum annealing against classical
algorithms (e.g., \cite{Hen:2015rt,2016arXiv160401746M,DW2000Q,Hamerly:2019aa,Junger:2019aa}). The TTS metric gives accurate scaling with problem size only when $t_\mathrm{f}$ is optimized
to minimize the TTS for each size~\cite{speedup,Albash:2017aa}. In our
experiments, we used a fixed $t_\mathrm{f}=5\ \mu$s, and hence these results only place a lower
bound on the true scaling~\cite{Hen:2015rt}, but this is sufficient for our purposes. Since our anneal time was fixed, we actually report the number of runs $R=\mathrm{TTS}/t_\mathrm{f}$, which we still refer to as TTS. Note that $R\geq1$ as one needs to run the annealer at least once to find
the correct ground state.

In~\cref{fig:rts_noise}, we show the TTS required to find the ground state for the
median instances at each size, for both C and QAC, sorted by different levels of added noise $\eta$. As expected for spin glasses, the TTS scales at least exponentially in
$L$ for both cases, with the scaling becoming worse for larger $\eta$. However, we note that QAC exhibits both milder scaling and lower absolute effort.
The same conclusion holds when we sort instances by hardness (see~\cref{app:hardness}). To more directly see the advantage of QAC over C, we plot the speedup ratio~\cite{Hen:2015rt} in~\cref{fig:speedup}. This plot clearly shows the scaling advantage of QAC over C for sufficiently large size $L$ and added noise $\eta$: for all added noise levels $\eta$, the slope of the speedup ratio becomes positive beyond an initial transient at small sizes $L$, and this happens sooner the larger $\eta$ is.

\subsection{Data collapse and scaling: doom \textit{vs} hope}

We have seen that QAC outperforms the C strategy. But what is the worst-case classical cost of solving the same Ising problem instances? For a generalized Chimera graph of $L \times L$ unit cells of complete bipartite graphs $K_{r,r}$, the tree-width is $w=rL+1$; for the D-Wave devices used here $r=4$. Dynamic programming takes time $O(L^2 2^w)$ to find the ground state of any Ising problem defined on such a graph~\cite{Selby:2014tx,Junger:2019aa}. Here $2^w$ is the dimension of the exhaustive search space for each of the $L^2$ tree nodes of width $w$. Thus in the present case any problem can be solved exactly, deterministically, in time TTS$_{\mathrm{DP}} = O(L^2 2^{4L})$.
However, adding analog errors exponentially suppresses the probability of success. Specifically, if the errors are drawn from a Gaussian distribution with standard deviation $\eta$ on an instance with $N$ spins, then $P_{g} = O(e^{-\eta^{\alpha} N})$, where $\alpha \leq 1$ depends on the problem class~\cite{albash_2019_analogerrors}. Thus, to find the ground state of the intended Hamiltonian $H_{\mathrm{Ising}}$, running dynamic programming on the Ising instances with noise added is expected to take a time scaling as $\mathrm{TTS}_{\mathrm{DP}} \times \left\lceil \frac{\ln(1-0.99)}{\ln(1-P_{g})} \right\rceil$, which reduces, in the limit $P_{g} \ll 1$, to:
\beq
 \mathrm{TTS}_{\mathrm{DP}}/P_{g}   =  O(L^{2} 2^{4L} e^{c_{\mathrm{DP}} L^{2}})  \ , \ \ c_{\mathrm{DP}}=8 \eta^{\alpha}
\label{eq:DP}
\eeq
since the dynamic programming algorithm is only presented the intended Ising instance once every $1/P_g$ times on average. Thus the worst case classical cost is asymptotically $O(e^{8 \eta^{\alpha} L^2})$. Note that this scaling is determined not by the intrinsic performance of the DP algorithm, but by the probability $P_g$ that it is presented with the intended Hamiltonian, which is algorithm-independent (and is, of course, not a problem an algorithm running on classical digital computers would need to suffer from). A random guess would find the intended ground state with the same asymptotic scaling: $ \mathrm{TTS}_{\mathrm{rand}}/P_{g} = 2^{N} e^{8 \eta^{\alpha} L^{2}} = O(e^{c_{\mathrm{rand}} L^{2}})$, but with a larger exponent: $c_{\mathrm{rand}} = c_{\mathrm{DP}}+8\ln2$.

\begin{figure}[t]
  \centering
  \centering
  {\includegraphics[width=\columnwidth]
    {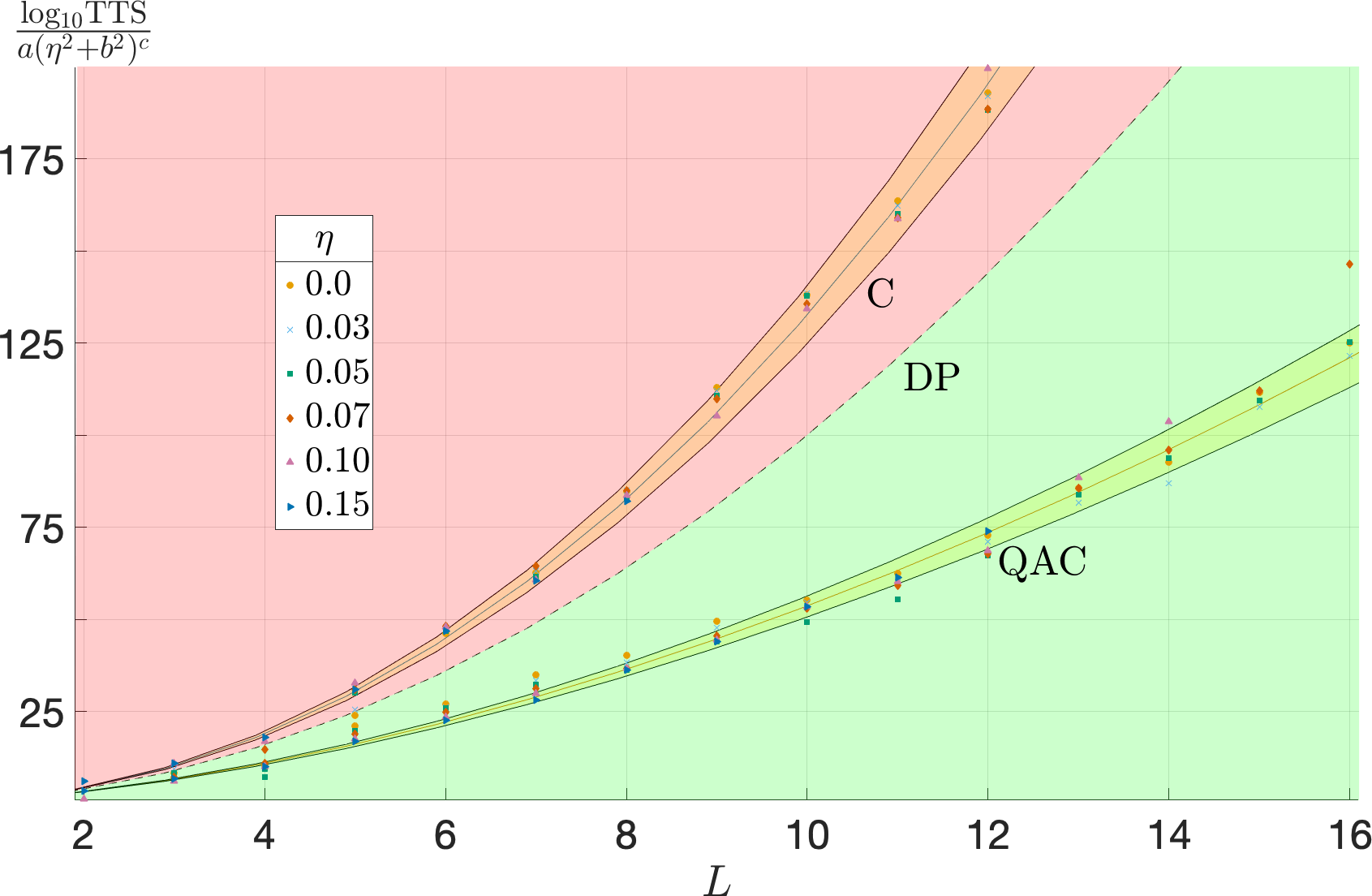} }
  \caption{\textbf{Data collapse and fit.} Result of the collapse of the
    data after fitting the TTS results shown in~\cref{fig:rts_noise} to $10^{a(\eta^{2}+b^{2})^{c}L^{d}}$ [\cref{eq:best-fit-function}]. The dashed line is the asymptotic scaling of the classical dynamic programming algorithm, for which $\log_{10}(\mathrm{TTS}) \sim L^2$. The red/green region above/below this line is where the C/QAC data lies after the data collapse. These correspond, respectively, to a guaranteed slowdown and a possible (but not guaranteed) speedup. The blue and red solid are the fits derived from the data collapse, with parameters given in~\cref{tab:fits}. The shaded regions around the fitted lines represent the $95\%$ C.I. fits as described in~\cref{sec:appendix_scaling}.}
  \label{fig:finite_scaling}
\end{figure}

To compare the D-Wave device's TTS scaling to this form, we attempted a data collapse of the results shown in~\cref{fig:rts_noise}, in order to include the $\eta$ dependence in the scaling function. To achieve the data collapse we ran a comprehensive search for functions $f(L,\eta)$ that would collapse both the C and QAC data using as few fitting parameters as possible  (see Methods~\cref{sec:finite_scaling} for details of the procedure). A natural choice for such a function is a generalization of Eq.~\eqref{eq:DP} with up to five free parameters, of the form $\mathrm{TTS} = a L^{2} 10^{bL+c (\eta^2 + d^2)^e L^{2}}$. However, we found it to perform poorly (see~\cref{sec:appendix_scaling}). Instead, we found that the four-parameter form
\beq
f(L,\eta) = 10^{a(\eta^2 + b^2)^c L^d} ,
\label{eq:best-fit-function}
\eeq
where the crucial difference is the replacement of $L^2$ by $L^d$ in the exponential,
works very well for both the C and QAC data (using three or fewer parameters gives poor agreement).
The data collapse and fit results are shown in~\cref{fig:finite_scaling}, and the fit parameters along with their $95\%$ C.I. are given in~\cref{tab:fits}. The relatively tight error bounds are evidence of the quality of the data collapse.

Surprisingly, we find that $d>2$ for the C strategy, with high statistical confidence. This means that without error suppression, and even after using a majority vote among four copies of the problem, the performance of the quantum annealer is worse than that of a deterministic worst-case classical algorithm, for which $d=2$. Hence the ``doom'' advertised in the title of this work.

Fortunately, not all is lost: this disturbing finding is mitigated by QAC. As seen in~\cref{fig:finite_scaling} and~\cref{tab:fits}, for QAC we obtain  $d<2$, again with high statistical confidence. This result restores the hope that a quantum annealer can eventually become competitive with classical optimization algorithms, but only after the incorporation of an error suppression and correction strategy such as QAC.

\begin{table}[t]
\begin{center}
\begin{tabular}{|c|c|c|c|c|}
\hline
Scheme & $a$ & $b$ & $c$ & $d$\\
\hline
\textbf{C} & \textbf{8.01} & \textbf{0.134} & \textbf{1.61} & \textbf{2.12}\\
\hline
C lower & 5.71 & 0.109 & 1.58 & 2.10 \\
\hline
C upper & 10.3  &  0.159 & 1.64  & 2.15 \\

\hline
\textbf{QAC} & \textbf{0.392} & \textbf{0.069} & \textbf{0.486} & \textbf{1.73} \\
\hline
QAC lower &  0.384 & 0.057 & 0.483  & 1.70 \\
\hline
QAC upper & 0.399 & 0.081 & 0.490 & 1.75 \\
\hline
\end{tabular}
\end{center}
\caption{Fit parameters for Eq.~\eqref{eq:best-fit-function} after data collapse of the TTS scaling data shown in~\cref{fig:rts_noise}. ``Upper'' and ``lower'' refers to the $95\%$ C.I. values of the parameters, calculated as explained in detail in~\cref{sec:appendix_scaling}. Of particular note is the $d$ parameter, which determines the asymptotic scaling. For QAC $d<2$ while for C $d>2$, with $d=2$ being the scaling of an exhaustive classical solver.}
\label{tab:fits}
\end{table}%

\section{Discussion}
\label{sec:discussion}

It should be remarked that our results on optimization have no direct bearing on other tasks quantum annealers are potentially capable of speeding up, such as approximate optimization~\cite{King:2015cs,Vinci:2016tg} and sampling~\cite{2012arXiv1204.2821S,Amin:2016,Mott:2017aa,Li:comp-bio-2017,Perdomo-Ortiz:2018aa}. Nor do our results address quantum annealing slowdowns due to small gaps~\cite{vanDam:01,Reichardt:2004,Jorg:2010qa,Laumann:2012hs}, which may be addressed via other methods, such as non-stoquastic Hamiltonians~\cite{Nishimori:2016aa,Albash:2019aa}, reverse annealing~\cite{Perdomo-Ortiz:2011fh,Chancellor:2016ys,Ohkuwa:2018aa}, or inhomogeneous transverse field driving~\cite{Susa:2018aa,Susa:2018ab}. However, none of these methods is immune to the effects of $J$-chaos.

To conclude, we have shown that QAC can reduce the detrimental effects of $J$-chaos on the performance of quantum annealers. In the regime we tested, QAC becomes more effective the higher the noise is and the larger the problem size is. The improvements seen are distinctly greater than without error suppression and correction, even after equalizing resources in terms of total qubit count, in terms of both scaling and absolute effort. Moreover, QAC undoes a catastrophic loss to an exhaustive classical algorithm by improving the scaling of the annealer's TTS to below the classical upper bound. Thus, we have demonstrated that QAC is not only an effective tool that can be used to improve current quantum annealing hardware, but that error suppression and correction are essential to ensure competitive performance against classical alternatives. Further improvements using more powerful error suppression and correction strategies than the simple one we explored here are certainly expected, and undoubtedly necessary, as ultimately only a fully fault-tolerant approach is expected to be effective in the asymptotic limit of large problem sizes.

\part{}

\section*{Methods}
\label{sec:methods}

\subsection{Random Ising Instances}
\label{sec:insts}

\textbf{Without noise} --- The set of instances used were generated randomly on the \pudenzcode graph produced by the $L \times L$ Chimera graph for $L \in \{2,\dots,12\}$ on the D-Wave 2X and $L \in \{13,\dots,16\}$ on the D-Wave 2000Q. There were $100$ instances at each graph size such that the local fields, $h_i$, were $0$ and the couplings $J_{ij}$ were drawn uniformly at random from the set $\pm \frac{1}{6} \times \{1, 2, 3\}$. We found the ground state energy of these logical instances via the Hamze-Freitas-Selby (HFS) algorithm~\cite{Selby:2014tx,hamze:04} and parallel tempering with iso-energetic cluster (Houdayer) moves (PTICM)~\cite{Houdayer:2001aa,PhysRevLett.115.077201}. By using both, we consistently found ground state energies lower than or just as low as those found by the D-Wave devices. In a few instances the latter found lower energies than HFS (one instance at $L = 15$ and five at $L = 16$), and these were confirmed as correct using PTICM.
As such, we are confident that the ground state energies found were in fact correct.

\textbf{With noise} --- We generated random numbers $\delta J_{ij} \sim \mc{N}(0,\eta^2)$. If a modified coupler value $\tilde{J}_{ij} = J_{ij} + \delta J_{ij}$ fell outside the experimentally allowed range $[-1,1]$, we truncated it to $\pm 1$. Since the largest
coupling in our set was $0.5$ and the largest noise had a standard deviation of $\eta=0.15$,
these truncated values were only used a handful of times in our entire data collection.

\subsection{The D-Wave devices used}
\label{sec:d-wave}

\begin{figure*}
  \centering
  \subfigure[\ ]{\ig[width=0.15\textwidth]
    {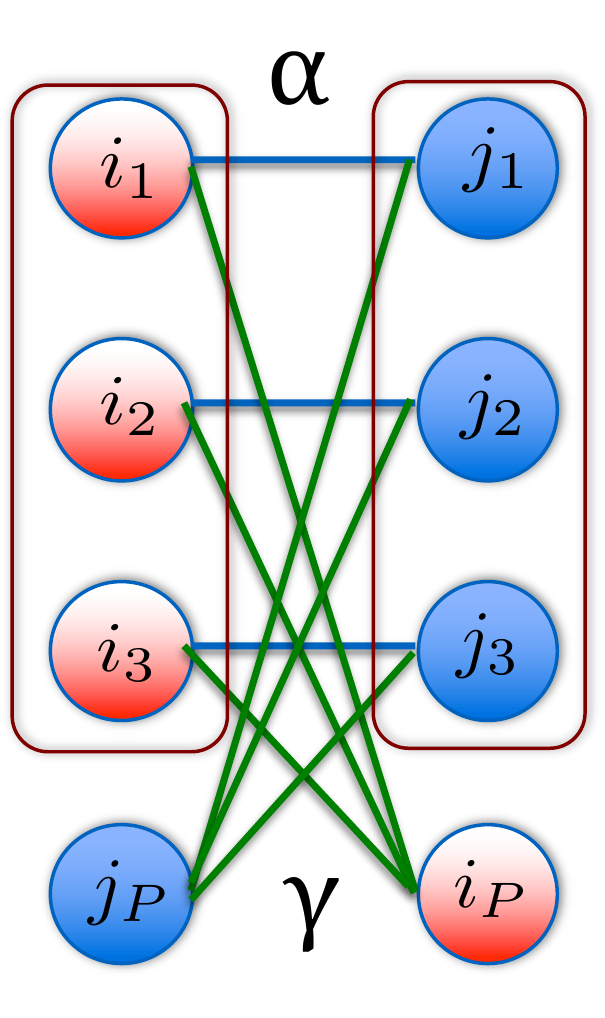}\label{fig:pudenz_code_unit_cell}}
  \subfigure[\ ]{\includegraphics[width=0.4\textwidth]
    {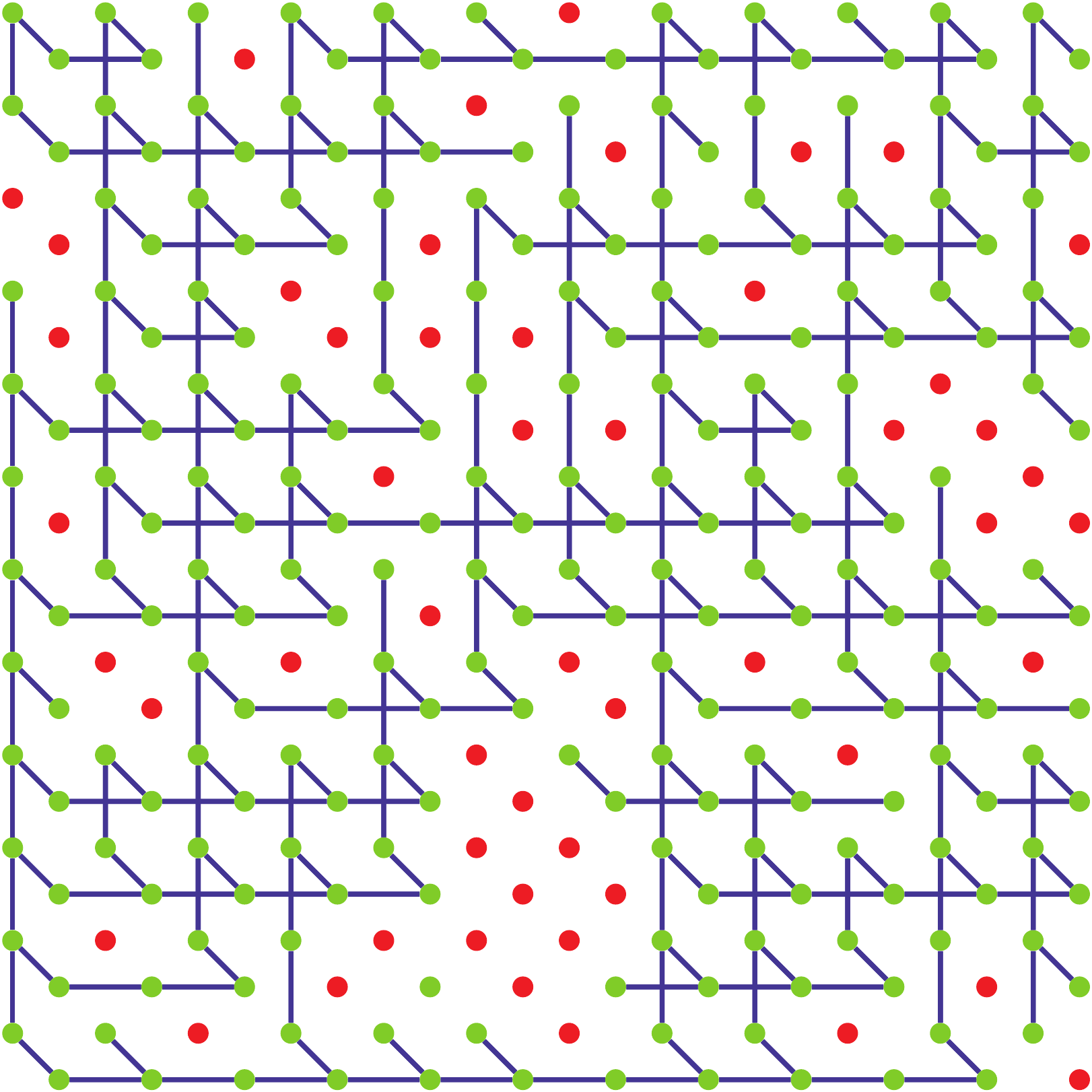} \label{fig:pudenzcode}}
  \subfigure[\ ]{\includegraphics[width=0.4\textwidth]
    {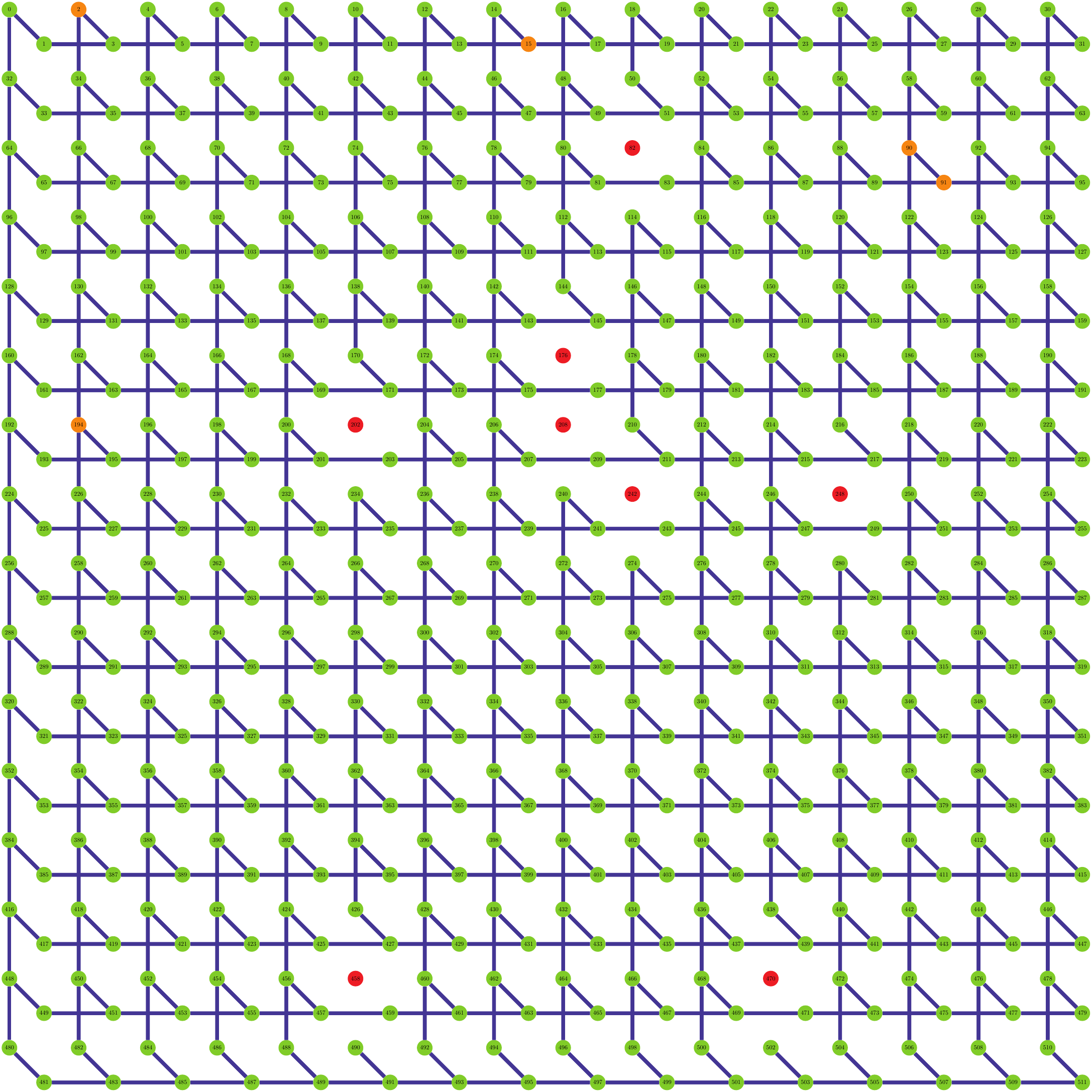} \label{fig:pudenzcode2000Q}}
  \caption{
  \textbf{The \pudenzcode and the resulting logical graphs of the DW2X and DW2000Q devices.}
  (a) A logical qubit of the \pudenzcode is formed by connecting
three qubits in one half of the unit cell to a single qubit on the other half of the unit
cell. The three qubits are called the ``data qubits'' and the single qubit plays the role
of a ``penalty qubit''. The data qubits are tied together with the penalty qubit by
ferromagnetic couplings to ensure consistent evolution of each of the three data
qubits. Each unit cell now contains two logical qubits. (b) and (c) Each logical qubit connects to its
neighbor in the same cell via the intra-cell couplings and connects to its horizontal or
vertical neighbor via the inter-cell couplings. This construction gives rise to a graph of degree
$3$ which is a minor embedding of the Chimera graph. (b) is the logical graph for the DW2X, (c) for the DW2000Q. In
    both graphs, green (red) circle denote operational (inactive) logical qubits. Orange circles denote logical qubits with intact data qubits but an inactive penalty qubit. The lines denote
    the active couplings between the operational logical qubits. For each problem size $L$, we used a subgraph formed by taking an $L\times L$ square starting from the top left corner.}
    \label{fig:logical-graphs}
\end{figure*}

In this study, we used the D-Wave 2X (DW2X) annealer installed at the USC Information
Sciences Institute and the D-Wave 2000Q (DW2000Q) annealer installed at the NASA Quantum Artificial Intelligence Laboratory (QuAIL). The qubits of the annealer occupy the vertices of the Chimera graph of size $12\times12$ for the DW2X and $16\times16$ for the DW2000Q (see~\cref{fig:graphs} in~\cref{sec:graphs}). The DW2X has $1098$ functional qubits, leading to a \pudenzcode graph with $236$ functional logical qubits and the DW2000Q has $2031$ functional qubits, leading to a \pudenzcode graph with $504$ functional logical qubits, as shown in Fig.~\ref{fig:logical-graphs}.

The annealing time $t_{f}$ can be chosen in the range $[5,2000] \mu$s on the DW2X and $[1,2000] \mu$s on the DW2000Q. We used $t_{f}=5 \mu$s, since this is the fastest time that could be used across both devices.

There were some differences between the structure and performance of these two devices. A discussion of these differences can be found in~\cref{app:difference}.

\subsection{Data analysis}
\label{sec:data-analysis}

We used a Bayesian bootstrap~\cite{rubin_1981_bayesianbootstrap} over the underlying data
(collected as described in Methods~\cref{sec:insts}) to compute the mean $\mu$ and the standard deviation $\sigma$ of the success probabilities and their associated error bars. The ground state probability $P_g$ used in~\cref{eq:8} is the same as $\mu$.

Consider $g$ gauges where for each gauge $i$ we find $s_{i}$ successful readouts out of
the total $M$ readouts. We used the Beta function $\beta(s_{i},M-s_{i})$ as our posterior
probability distribution of success, i.e., it is our best guess of the distribution of the
success probability, given the observation of $s_i$ successful hits in $M$ attempts. To
draw one sample of our bootstrap distribution, we did the following:

\begin{enumerate}[leftmargin=*]
\item First, sample from each of the $\beta$-distributions. Let $B_{i}$ be a sample from
  the distribution $\beta(s_{i},M-s_{i})$ and let $\vec{B}=\{B_{1},B_{2},\ldots,B_{g}\}$.
\item Then, sample a point from the $g$-dimensional uniform Dirichlet distribution. This
  is a $g$-dimensional vector $\vec{D}$.
\item The estimate of the success probability for this bootstrap sample is given by
  \begin{equation}
    \label{eq:11}
    \mu = \vec{D}\cdot \vec{B} = \sum_{i=1}^{g}D_{i}B_{i}
  \end{equation}
  and the estimate of the standard deviation of this sample is the square root of the
  weighted variance of $\vec{B}$ where the weights are given by $\vec{D}$,
  \begin{equation}
    \label{eq:12}
    \sigma^{2} = \sum_{i=1}^{g}D_{i}(B_{i}-\mu)^{2}\ .
  \end{equation}
  In~\cref{eq:11,eq:12}, we have used the fact that the samples of the uniform Dirichlet
  distribution sum to $1$: $\sum_{i}D_{i}=1$.
\item From these two quantities, we computed $\sigma/\mu$. Other quantities of interest can be
  similarly derived from a combination of $\vec{B}$ and $\vec{D}$.
\end{enumerate}

We repeated these steps a large number of times to obtain a bootstrap distribution over our
quantity of interest (in this case, $\mu$ and $\sigma/\mu$). Our best estimate of the quantity and
its associated error bars are given by the mean and the spread of the bootstrap
distribution respectively.

Data for C was collected as follows. For each instance and each noise value, we ran the annealer
with $5$ random gauges~\cite{q-sig,Job:2017aa} with $10,000$ readouts each. If $p$ is the success probability of
one unencoded copy and each copy is statistically independent,
then the C strategy will have at least one successful copy with probability
$1-(1-p)^{4}$. Here, we only collected data for a single copy, and then
used this combinatorial formula to get an estimate on the success probability of the
classical repetition case, as in earlier work~\cite{PAL:14}. In fact, due to crosstalk this provides an upper bound on the actual performance of the C strategy (see~\cref{sec:eta0}), so that our results favor QAC over C even more than our plots indicate.

Data for QAC (see~\cref{sec:QAC}) was collected as follows. Every data collection run for problem instance
$i$ of size $L$ can be labeled by two additional parameters; the strength of artificial
injected noise $\eta$ and the strength of the penalty value $\gamma$. The penalty strength was chosen
from the set $\gamma \in \{0.1,0.2,\ldots,0.5\}$. For each data collection run
$(i,L,\eta,\gamma)$, we ran the annealer with $5$ random gauges with $10,000$ readouts each, for a
total of $50,000$ readouts. From this data we estimated the mean success probability over
the gauges, $P_g(i,L,\eta,\gamma)$. The optimal penalty value
\begin{equation}
  \label{eq:10}
  \gamma_{\mathrm{opt}} = \argmax_{\gamma} P_g(i,L,\eta,\gamma)
\end{equation}
maximizes the success probability within the chosen range of penalty values. The results shown in~\cref{sec:results} were picked to be at this optimal value for each instance. Histograms of the optimal strengths for each problem size are shown in~\cref{app:pen}.

\subsection{Quantum annealing correction}
\label{sec:QAC}

In QAC we encode each logical Pauli-Z operator as a sum of $n$ such physical operators, i.e.,
\begin{align}
\overline{\sigma_{i}^{z}} = \sum_{l=1}^{n} \sigma_{i_l}^{z}\ , \qquad
\overline{\sigma_{i}^{z}\sigma_{j}^{z}} = \sum_{l=1}^{n} \sigma_{i_l}^{z} \sigma_{j_l}^{z} .
\end{align}
Furthermore, we add a term to ferromagnetically couple the physical copies through an auxiliary qubit, i.e.,
\beq
H_{P} = - \sum_{i=1}^N \sigma_{i_p}^{z}\overline{\sigma_{i}^{z}} ,
\eeq
where $N$ is the number of logical variables in the original optimization problem.
We refer to the physical copies, $\sigma_{i_l}^{z}$, as data qubits and the auxiliary qubit, $\sigma_{i_p}^{z}$, as a penalty qubit. Thus, we arrive at the following encoding of our logical problem:
\beq
 \overline{H}(s) = A(s)H_X + B(s)(\alpha \overline{H}_{\mathrm{Ising}} + \gamma H_P) ,
 \eeq
where $\overline{H}_{\mathrm{Ising}}$ is the encoded version of ${H}_{\mathrm{Ising}}$ in Eq.~\eqref{eq:2}, i.e., ${\sigma_{i}^{z}} \mapsto \overline{\sigma_{i}^{z}}$ and ${\sigma_{i}^{z}\sigma_{j}^{z}} \mapsto \overline{\sigma_{i}^{z}\sigma_{j}^{z}}$, and $\alpha$ is an overall energy scale for the problem Hamiltonian (not used in this work, but complementary to adding control noise~\cite{vinci2015nested}).
When we add control noise to the QAC Hamiltonian, we replace $h_i$ by $\tilde{h}_{i} = h_{i} + \delta h_{i}$ and $J_{ij}$ by $\tilde{J}_{ij} = J_{ij} + \delta J_{ij}$, with the noise satisfying Eq.~\eqref{eq:deltas}.

The current generations of D-Wave devices allow a direct implementation of this code in the Ising Hamiltonian for $n = 3$, as shown in~\cref{fig:logical-graphs}, but are unable to encode the driver Hamiltonian $H_X$, as this requires many-body terms of the form $(\sigma^x)^{\otimes n}$. Thus, increasing the penalty strength, $\gamma$, begins to diminish the effect of the quantum fluctuations that drive quantum annealing. On the other hand, larger $\gamma$ values are more able to suppress bit flip errors. Thus, there exists an optimal value of $\gamma$ which depends on the spectrum of the problem
instance~\cite{PAL:13,PAL:14,Vinci:2015jt,Mishra:2015}. This optimization is further discussed in Methods~\cref{sec:insts}.

After annealing, we obtain a state vector where each data qubit is measured in the computational basis. From this, we can obtain a state vector of logical qubits via a variety of decoding strategies~\cite{Vinci:2015jt}. In this work, we exclusively used the strategy in which each logical qubit is decoded by a majority
vote of its constituent data qubits.

\subsection{Data collapse}
\label{sec:finite_scaling}

Here we explain our procedure for identifying the optimal fit and data collapse function, and for extracting confidence intervals (C.I.'s) and error bars.
We considered trial TTS functions of the form $f(L,\eta) = 10^{g_i(L,\eta)}$, with:
\bes
\label{eq:fitg}
\begin{align}
  g_1 &= a(\eta^2 + b^2)^c L^d\ , \ g_2 = g_1 +\log_{10}(e) \\
  g_3 &= a L + c (\eta^2 + b^2)^{d_1} L^{d_2}  + \log_{10}(eL^2) .
\end{align}
\ees
For $g_3$ we focused on the three cases $\{d_1=1/2,d_2=2\}$, $\{d_1=d,d_2=2\}$, $\{d_1=1/2,d_2=d\}$. Thus our trial functions had either four ($\{a,b,c,d\}$) or five ($\{a,b,c,d,e\}$)  free fitting parameters. For each trial function we computed non-linear least-square fits to the median TTS data for C on $L \in \{2,\dots ,12\}$, and QAC on $L \in \{2,\dots, 16\}$. The fitting parameters were initially allowed to take any values. However, we only accepted fits with $a\geq 0$ in $g_3$, since $a<0$ (scaling that decreases with $L$) would have to reflect overfitting. Thus, we also computed fits where we squared all the fitting parameters [i.e., replaced $a$ by $a^2$ in~\cref{eq:fitg}, etc.] in order to enforce positivity. Furthermore, we tested if the discrepancy between the ideal and actual number of Chimera graph couplers made a difference by fitting with an effective $L$; see~\cref{app:difference} for details. Thus, for each trial function there were four different methods: unconstrained/squared fitting parameters with $L$/effective $L$. Lastly, all fits were attempted with each of the optimization methods possible in Mathematica: SimulatedAnnealing, RandomSearch, NelderMead, and DifferentialEvolution.

Across the different methods and optimization algorithms used, $g_1$ was consistently the best of the $4$-parameter fits and was always very close to $g_2$, which is its $5$-parameter generalization. The $g_3$ functions always resulted either in $a<0$ or otherwise a very poor fit. Parameter squaring also improved the fit quality, and of the optimization methods only NelderMead tended to give inferior results.

After determining the three parameters $\{a,b,c\}$ for the median TTS data for $g_1$, we found least-squares fits to the upper and lower bounds determined by the $95\%$ C.I.'s for the median TTS, by using the same set $\{a,b,c\}$ and letting only $d$ be a free parameter. In this manner we found $d_-$ and $d_+$, the exponents that provide respective lower and upper bounds on $d$ for the median TTS data. In turn, $d_-$ and $d_+$ have associated $95\%$ C.I.'s, denoted $\Delta d_-$ and $\Delta d_+$. The reported range of $d$ in~\cref{tab:fits} is then $[d_- -\Delta d_-,d_+ + \Delta d_+]$. The resulting fits for each $\eta$ are shown in~\cref{sec:appendix_scaling}.

\section*{Data availability}
All raw data is available upon reasonable request from the authors.
A Mathematica notebook containing the TTS results, analysis scripts, error analysis, and our detailed fitting and data collapse results is available~\cite{datacollapse.nb}.


%

\acknowledgments
We are grateful to Tameem Albash for useful discussions. The computing resources used for this work were provided by the USC Center for High
Performance Computing and Communications. The research is based upon work (partially)
supported by the Office of the Director of National Intelligence (ODNI), Intelligence
Advanced Research Projects Activity (IARPA), via the U.S. Army Research Office contract
W911NF-17-C-0050. The views and conclusions contained herein are those of the authors and
should not be interpreted as necessarily representing the official policies or
endorsements, either expressed or implied, of the ODNI, IARPA, or the U.S. Government. The
U.S. Government is authorized to reproduce and distribute reprints for Governmental
purposes notwithstanding any copyright annotation thereon. This work was partially supported by Oracle.

\appendix
\begin{center}
\large{Appendix}
\end{center}

\section{Chimera graphs of the D-Wave devices used in this work}
  \label{sec:graphs}

\begin{figure}
  \centering
  \subfigure{\includegraphics[width=.71\columnwidth] {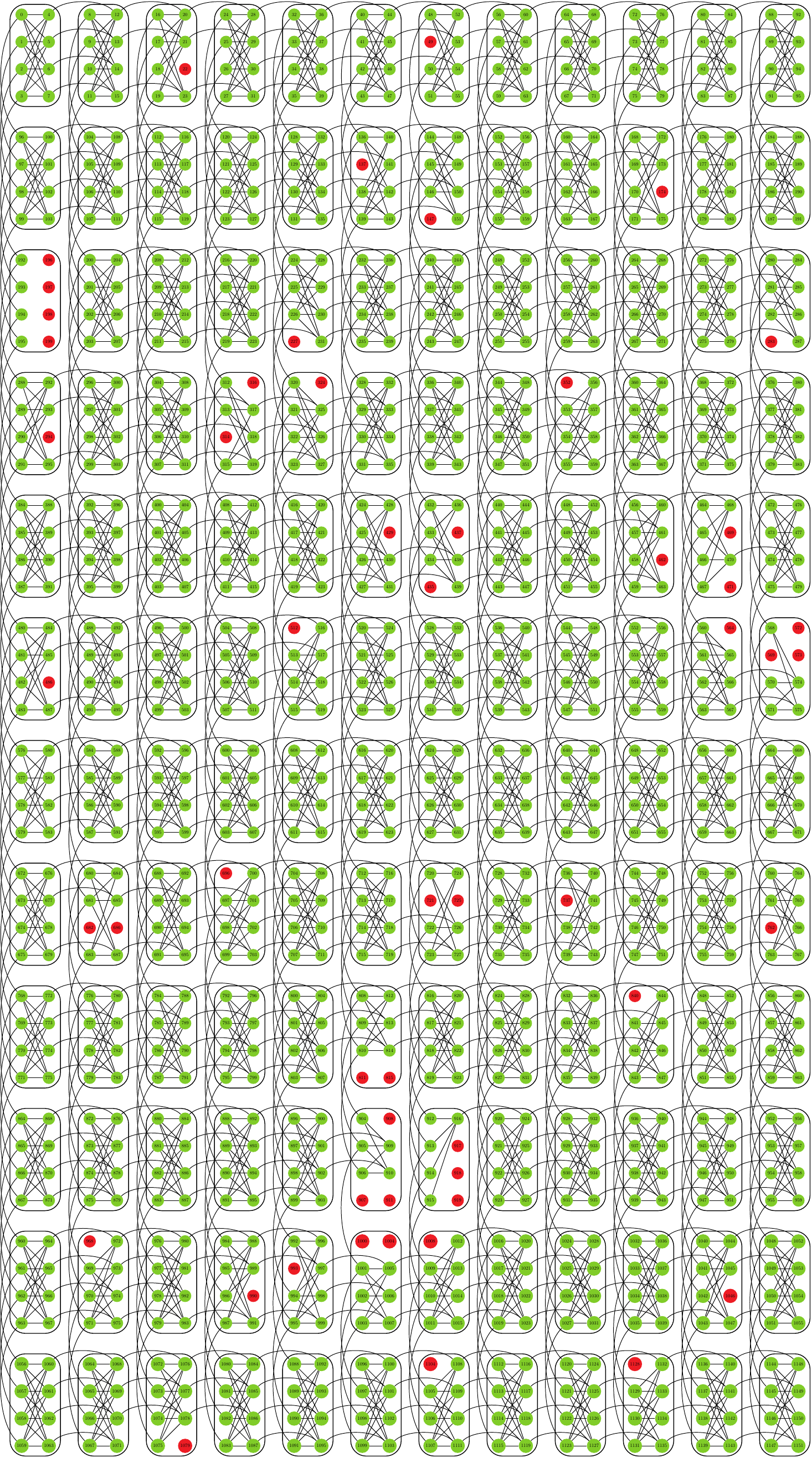} \label{fig:chimeragraph}}
  \subfigure{\includegraphics[width=.71\columnwidth] {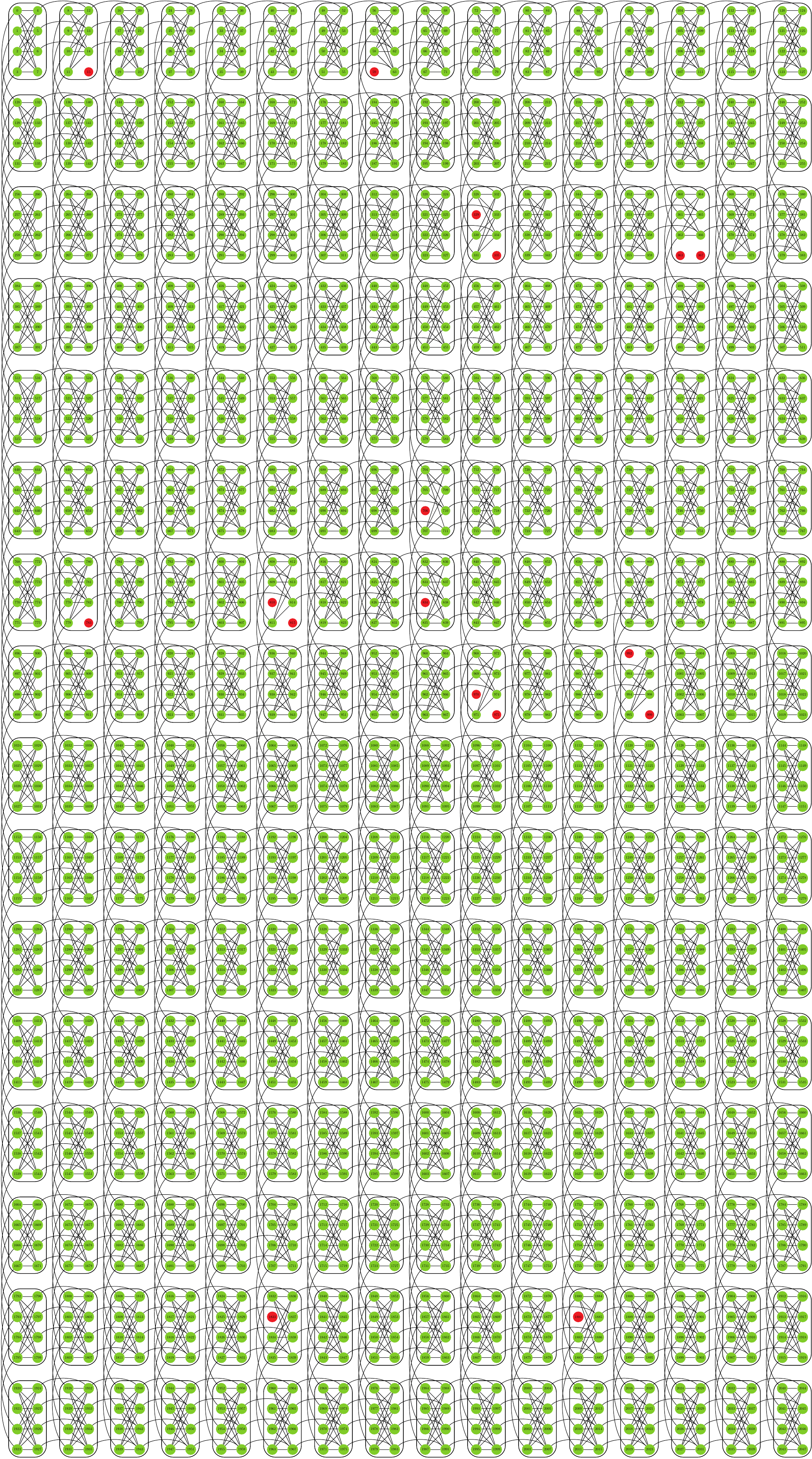} \label{fig:2000Qgraph}}
  \caption{\textbf{Chimera graphs.} The $12\times12\times8$ DW2X graph (top) and the $16\times16\times8$ DW2000Q graph (bottom).}
  \label{fig:graphs}
\end{figure}

The Chimera graphs of the DW2X and DW2000Q devices we used are shown in Fig.~\ref{fig:graphs}. In
    both graphs, green (red) circles denote operational (inactive) physical qubits. The lines denote
    the possible coupling between the operational physical qubits. Minor embedding of the \pudenz code leads to the logical graphs shown in Fig.~\ref{fig:logical-graphs}.

\section{Additional results}

\subsection{Fraction of failures of both QAC and C}
\label{app:fail}
\begin{figure}
  \centering
{\ig[width=\columnwidth]{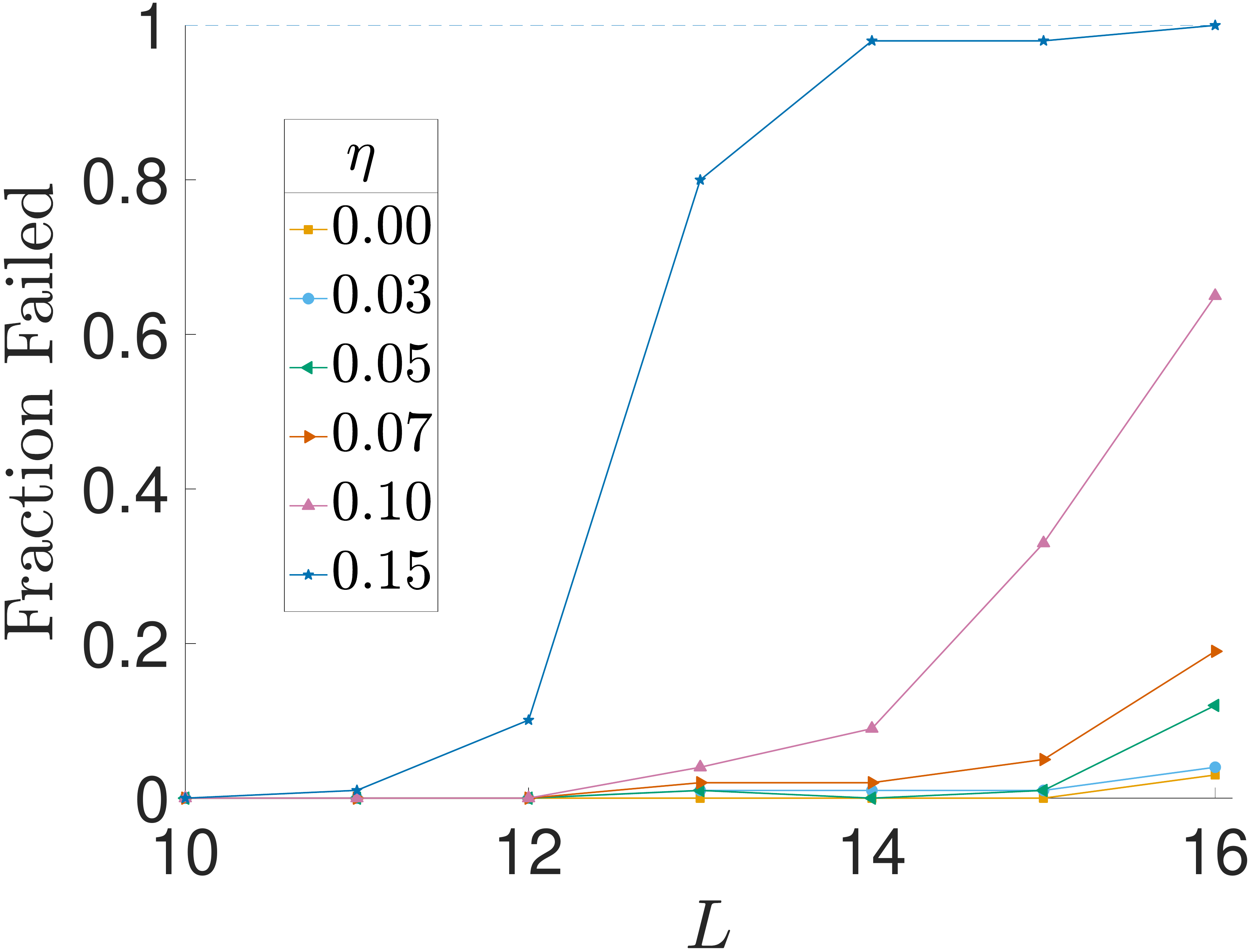}}
  \caption{\textbf{Fraction of instances where both C and QAC failed to find the ground state.} All instances are solved by either QAC or C for $L\leq 10$. The fraction of failures rises more rapidly as $\eta$ grows from $0$ to $0.15$, and only at the extreme of $L=16,\eta=0.15$ are no instances solved by either strategy.}
     \label{fig:failures}
\end{figure}

Success probability drops as more noise is added and problem size grows. \cref{fig:failures} shows the fraction of instances where neither QAC nor C found the ground state. This figure complements~\cref{fig:betters_success}, which includes all other instances, and shows that QAC improves upon C for sufficiently large values of $L$ and $\eta$.

\subsection{TTS Scaling with Size and Hardness}
\label{app:hardness}

\begin{figure}
  \centering
  \subfigure[\ C]{\ig[width=\columnwidth]
    {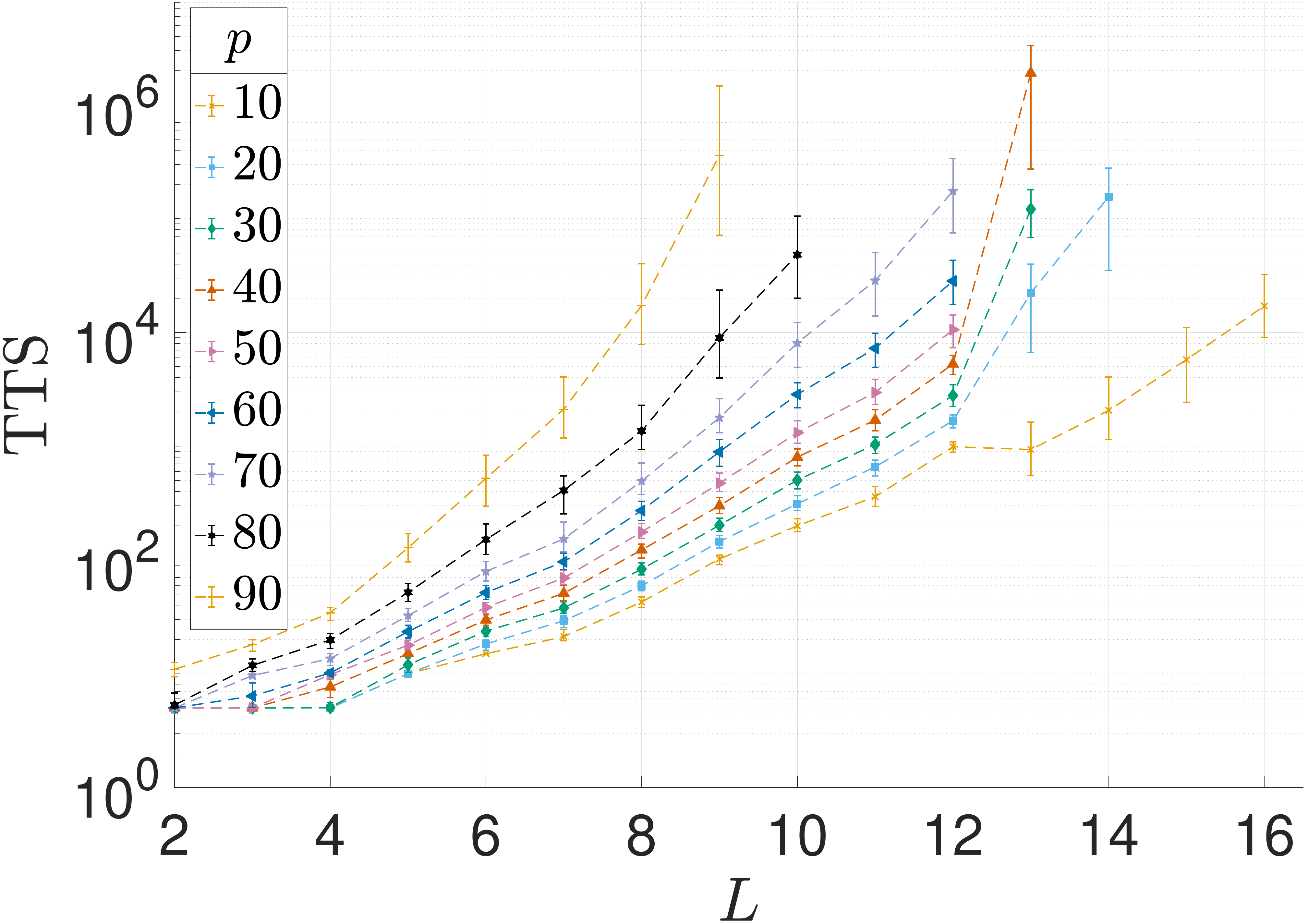}\label{fig:rts_percentile_C}}
  \subfigure[\ QAC]{\ig[width=\columnwidth]
    {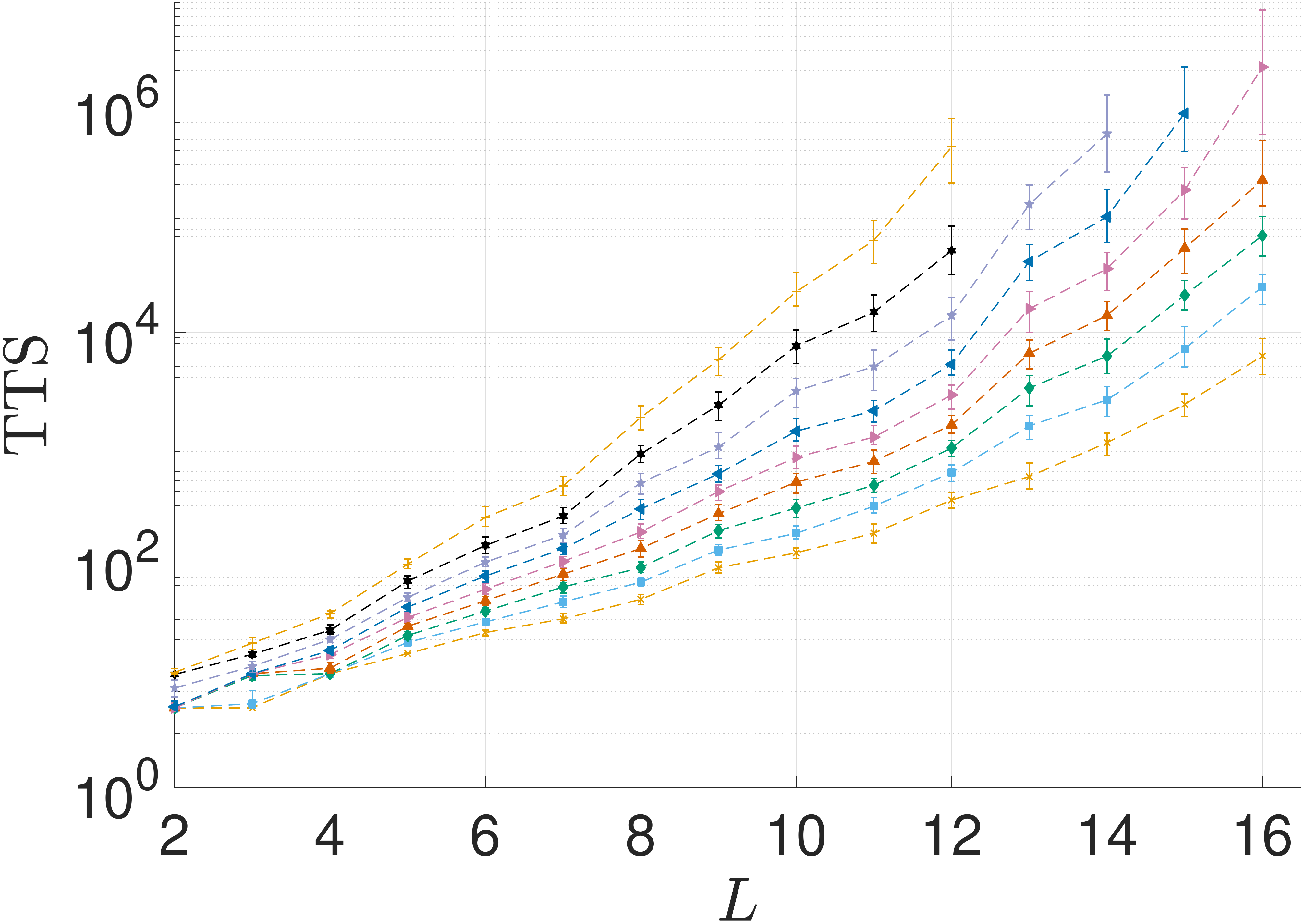}\label{fig:rts_percentile_QAC}}
  \caption{\textbf{TTS for QAC and the C strategy, sorted by hardness.} We show the number of runs
    $R$ required to find at least one ground state of the different
    hardness class of these instances. Here, for each size $L$, we group together the
    TTS of all the instances at every noise realization. From this group,
    we pick the $p^\mathrm{th}$-hardest percentile instance with $p\in\{10,\dots,90\}$. Then, we
    pick out the proper percentile instances, which now only depend on the system size
    $L$.  In (a), we show the results from the C strategy. In (b), we show
    the results for QAC. The scaling in (b) is milder compared to
    (a), demonstrating that QAC decreases the number of runs required in presence of
    $J$-chaos.}
     \label{fig:rts_percentile}
\end{figure}

\Cref{fig:rts_percentile} shows the same data as in~\cref{fig:rts_noise} when we combine all the noise realizations at each size $L$ and plot different percentiles of hardness. QAC improves the scaling at all percentiles, from the easiest instances at the $10^\mathrm{th}$ percentile to the hardest instances at the $90^\mathrm{th}$ percentile.

\subsection{Data collapse of TTS with error bars}
\label{sec:appendix_scaling}

\begin{figure*}
  \centering
  \subfigure[\ C]{\ig[width=2\columnwidth]
    {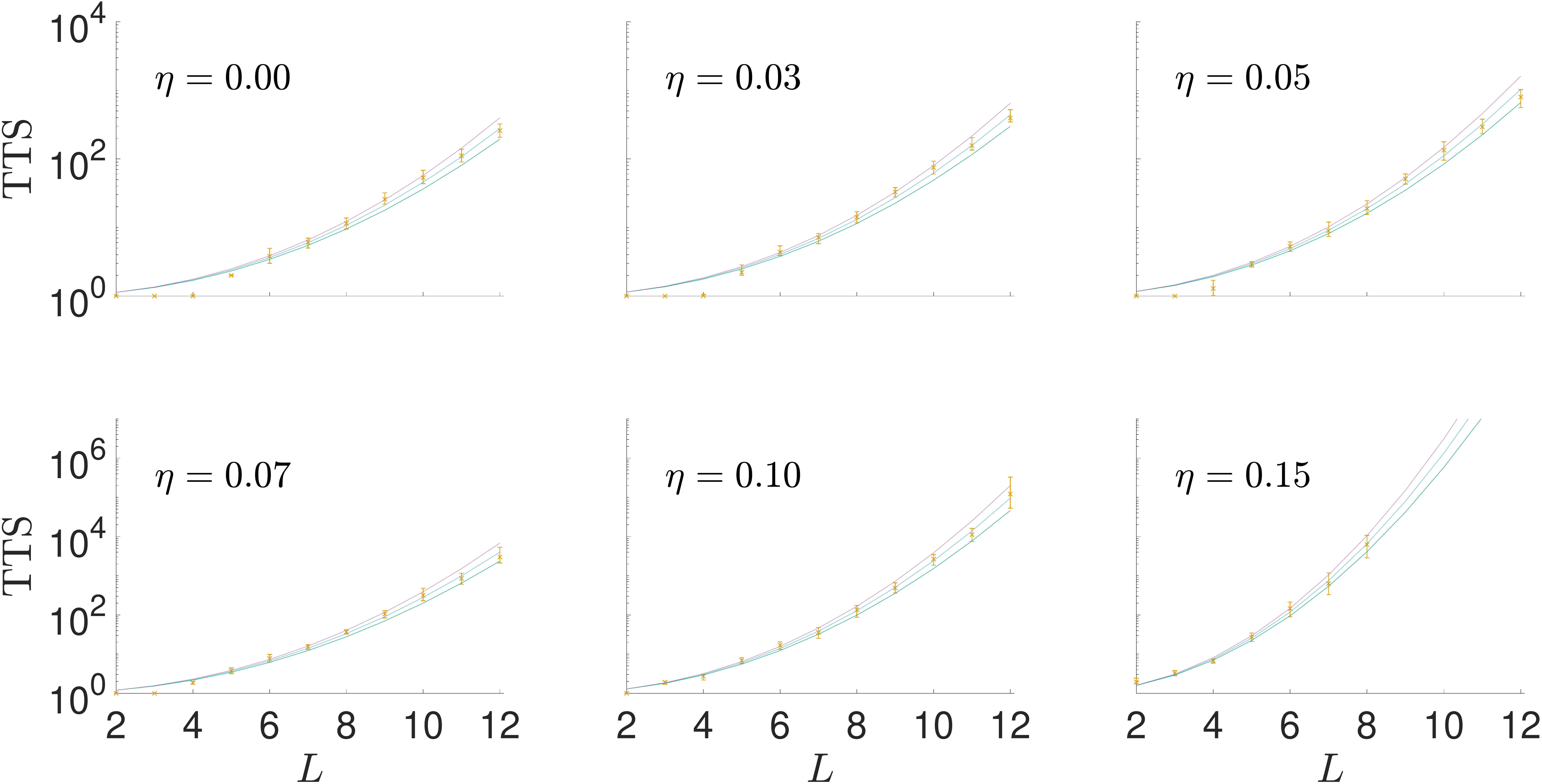}\label{fig:finite_scaling_C_1}}
  \subfigure[\ QAC]{\ig[width=2\columnwidth]
    {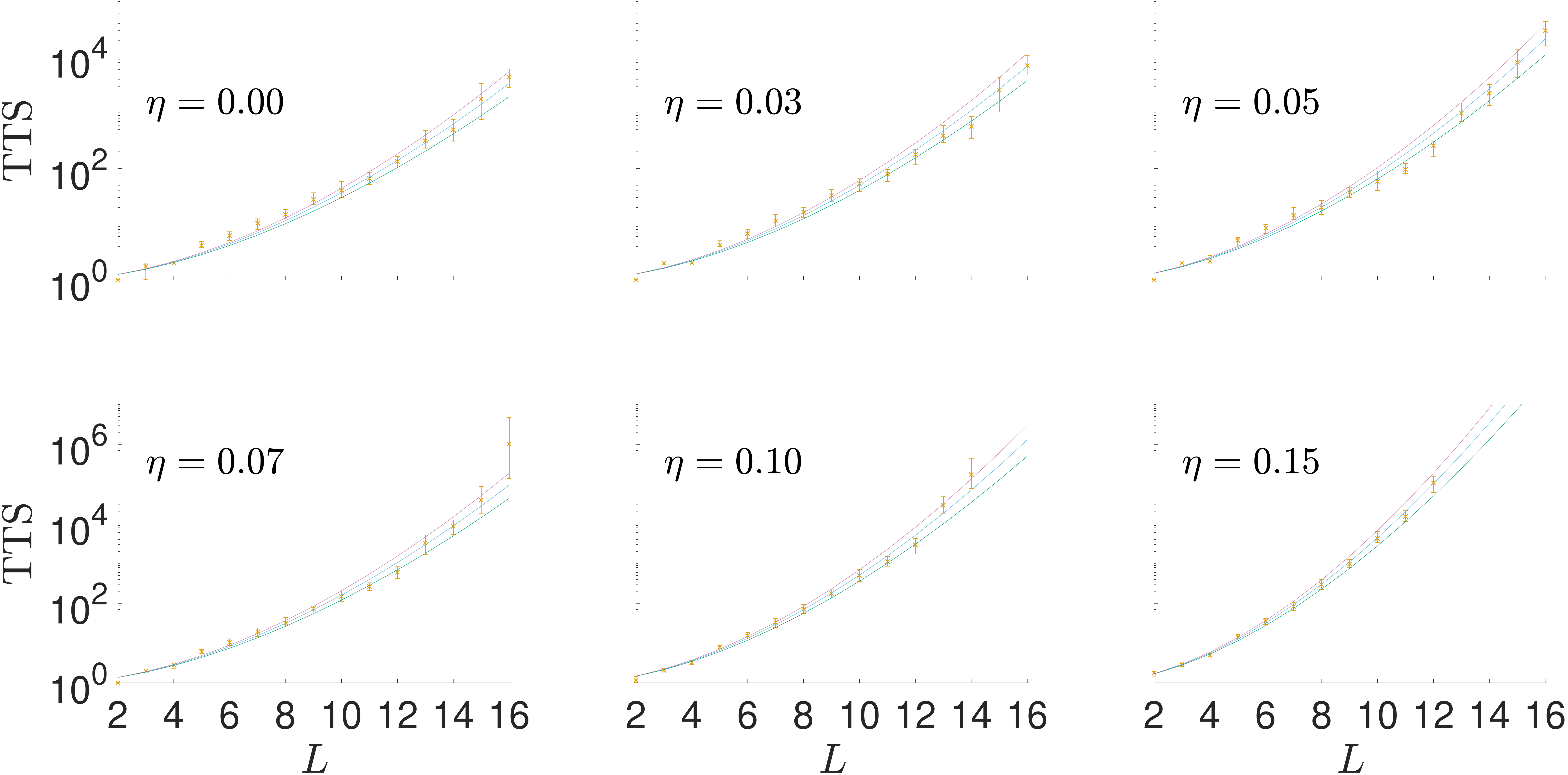}\label{fig:finite_scaling_QAC_1}}
  \caption{\textbf{Fits to the upper and lower error bounds of the median TTS data.} Error bars denote $95\%$ C.I.'s for the median TTS data (blue curves) at each value of added noise $\eta$. The fits shown are obtained by first performing a data collapse of the median data, then fitting $f(L,\eta) = 10^{a(\eta^2 + b^2)^c L^d} $ and extracting $\{a,b,c,d\}$, then using the obtained $\{a,b,c\}$ values to fit new exponents $d_+$ and $d_-$ for the upper and lower error bars, respectively. The resulting fits $f(L,\eta) $, with $d$ replaced by $d_{\pm}$, are shown in purple (upper bound) and green (lower bound).}
  \label{fig:collapse-with-errbars}
\end{figure*}

Figure~\ref{fig:collapse-with-errbars} shows the results of the fits computed for the data collapse procedure described in Methods~\cref{sec:finite_scaling}, including the error bars.

\section{Difference Between the DW2X and DW2000Q}
\label{app:difference}

\subsection{Coupler counts}
\label{sec:counts}

Since the DW2X and DW2000Q are different generations of the D-Wave devices, differing in both structure (see~\cref{fig:graphs}) and noise characteristics, performance differences are to be expected. In particular, the DW2X used has $1098$ functional qubits out of $1152$, and the DW2000Q used has $2031$ functional qubits out of $2048$. This leads to differences in the logical problems embeddable on each device, as we now explain in detail.

The couplings in the logical graphs of the \pudenzcode differ between the two devices, as seen in~\cref{fig:logical-graphs}.
Without any holes (missing physical qubits), in the logical graph each unit cell would be reduced to two logical qubits. These logical qubits would have one coupling between them, contributing a total of $L^2$ couplings to the logical problem. Furthermore, each unit cell would have one coupling to the unit cell below it, contributing another $L^2$, except that the last row of unit cells has no unit cells below it to connect to, so we have over-counted by $L$ couplings. The same analysis applies to the couplings to the right of each unit cell, contributing another $L^2 - L$ couplings. Thus, the total number of couplings in the ideal graph is $L^2 + 2(L^2 - L) = L(3L - 2)$. However, each hole in the physical graph contributes to the holes in the logical graph, removing some number of active couplers.

\begin{figure*}
  \subfigure[\ ]{\includegraphics[width=.99\columnwidth]{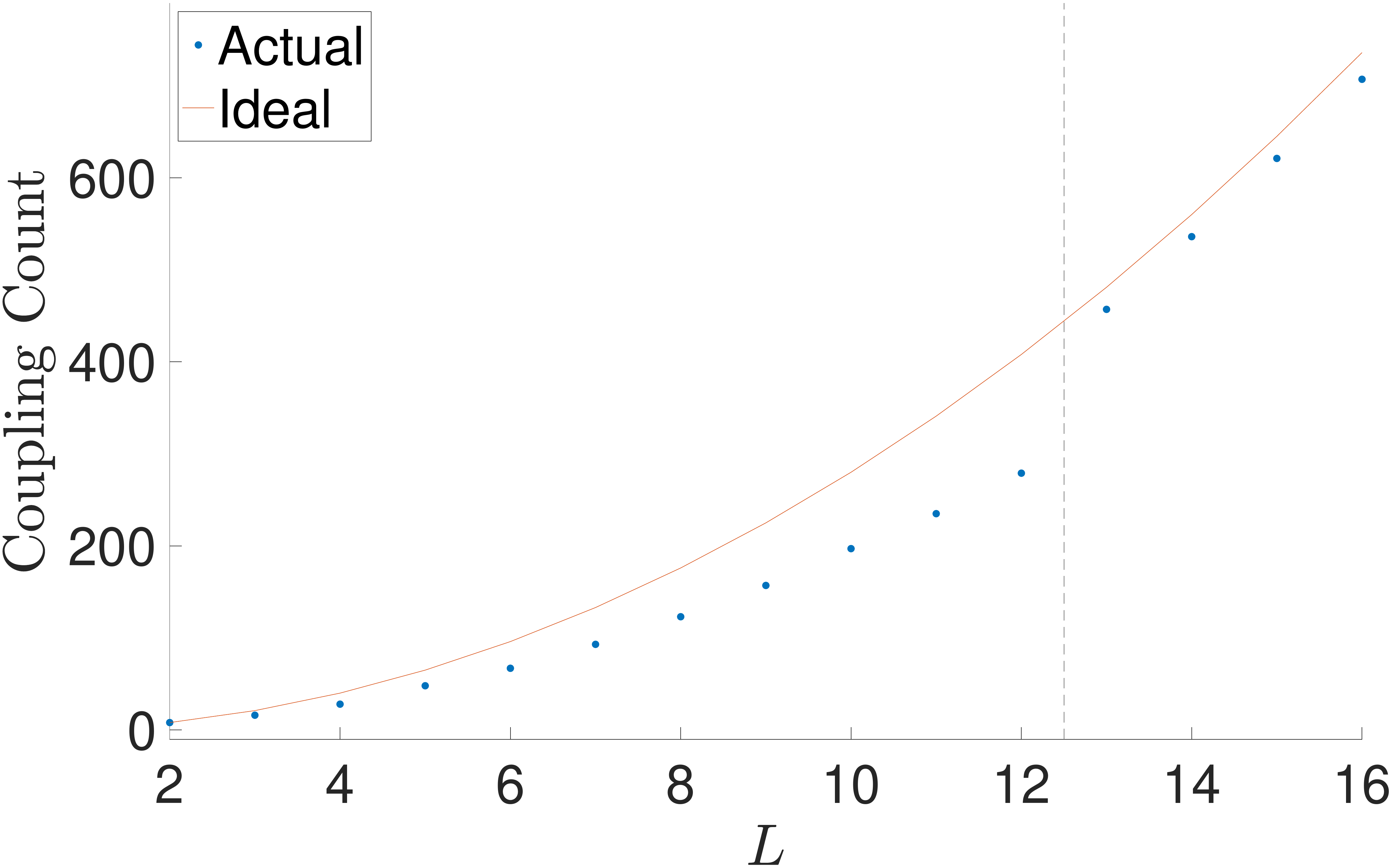}\label{fig:coupling_count}}
  \subfigure[\ ]{\includegraphics[width=.99\columnwidth]{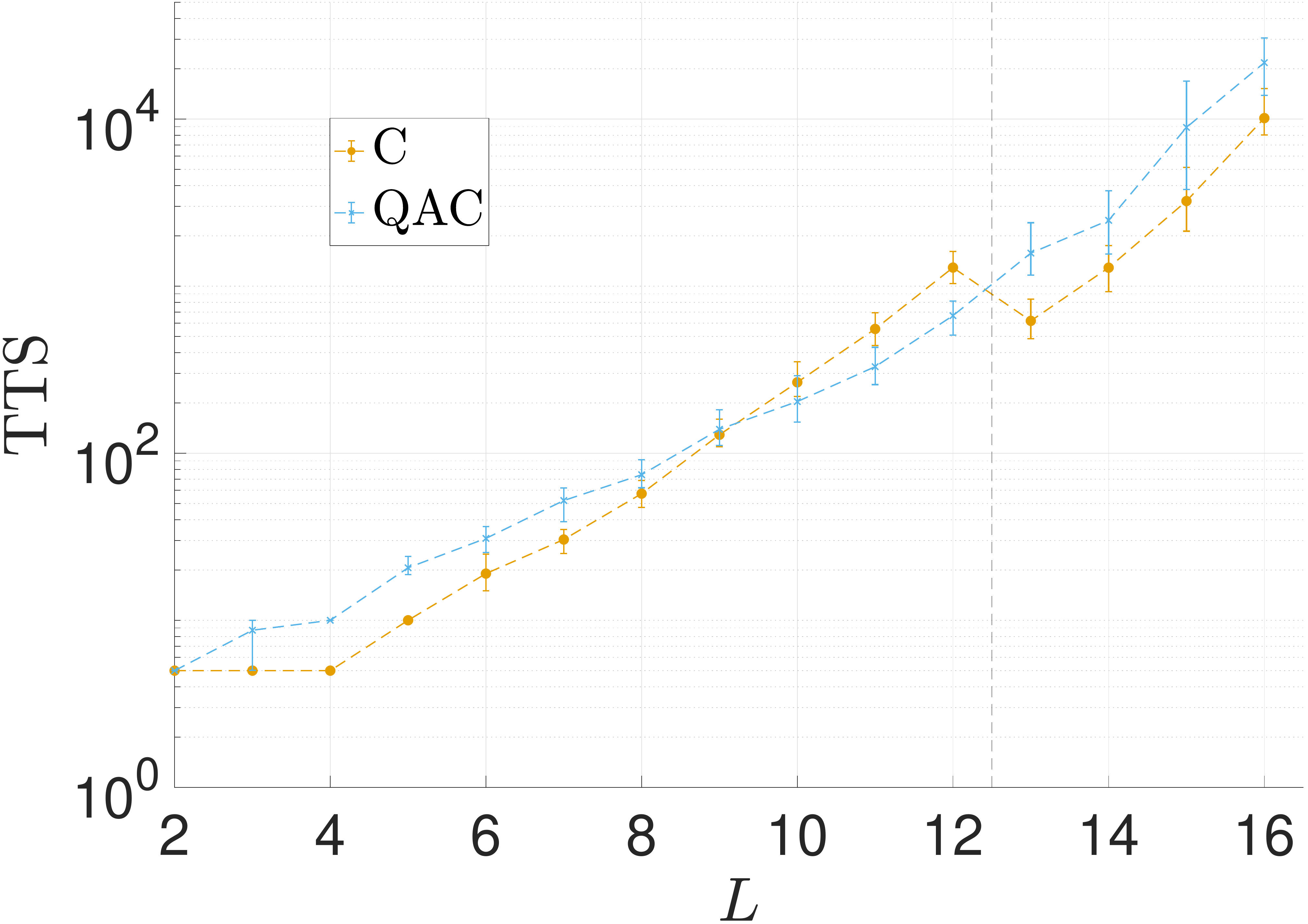}\label{fig:rts_noise_0}}
  \caption{\textbf{Effect of transition from DW2X to DW2000Q}. (a) Coupler counts in the logical graph of the \pudenzcode\!\!. We show the ideal number of couplers in a \pudenzcode graph with $L \times L$ unit cells, and the actual number in the two devices used, with the vertical dashed line marking the transition from the DW2X to the DW2000Q. (b) Runs-to-solution for QAC and C strategy for $\eta = 0$. We show the number of runs
    $R$ of the annealer required to find at least one ground state of the median instance
    of class $(L,\eta=0)$, for QAC and the C strategy. There is a sudden change in the C case for $L=13$, due to the transition from the DW2X to the DW2000Q device.}
  \label{fig:counts}
\end{figure*}

The difference between the ideal and actual number of couplers is shown in~\cref{fig:coupling_count}. As can be seen, there is a sudden jump from the DW2X to the DW2000Q in terms of number of couplers. Since the problem instances used in this work involve adding noise to each coupler present, this implies that at equal $(L,\eta)$ the problems solved on the DW2000Q are expected to be somewhat harder than on the DW2X.

\begin{figure}[t]
  \centering
  {\includegraphics[width=\columnwidth]
    {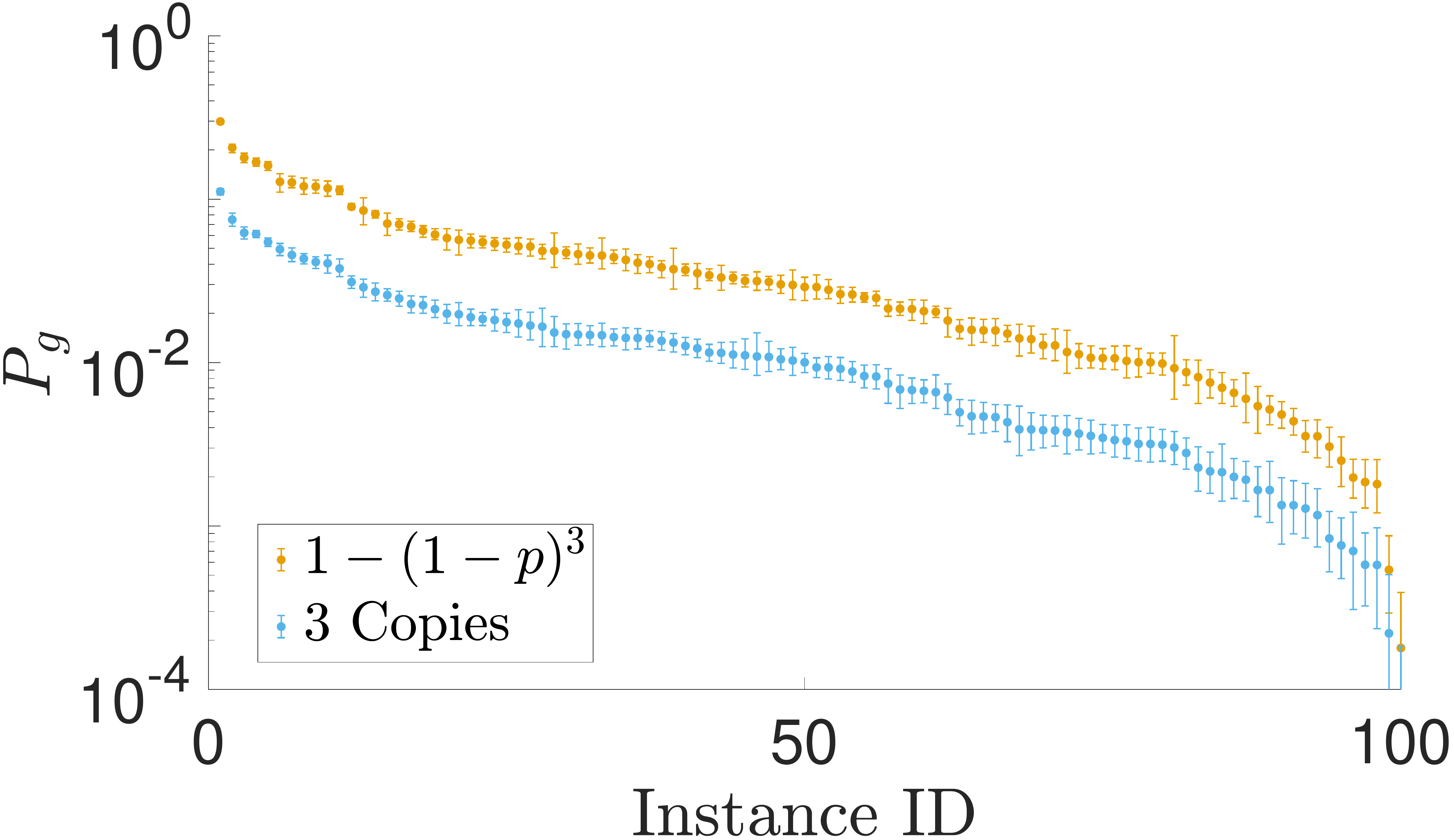}}
  \caption{\textbf{Cross Talk in Repetition Code.} We show the difference between the theoretical probability of finding the ground state at least once using three independent copies, and the actual result of using 3 physical copies of the same set of $100$ problem instances on the DW2X for $L = 13$ and $\eta=0$.}
  \label{fig:crosstalk}
\end{figure}

\begin{figure*}
  \centering
  \subfigure[\ $\eta=0.00$ ]{\ig[width=\columnwidth]
    {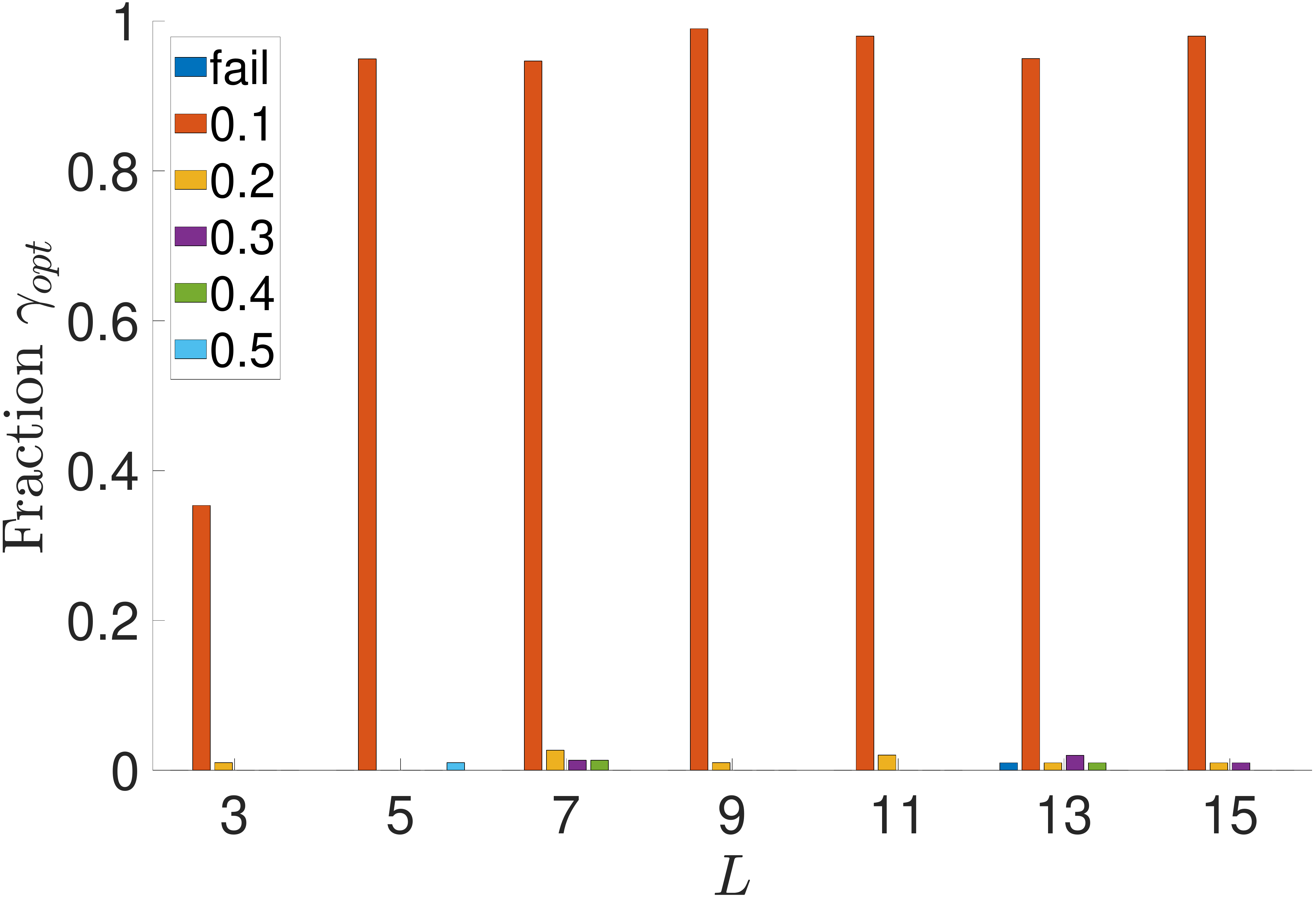}}
  \subfigure[\ $\eta=0.10$ ]{\ig[width=\columnwidth]
    {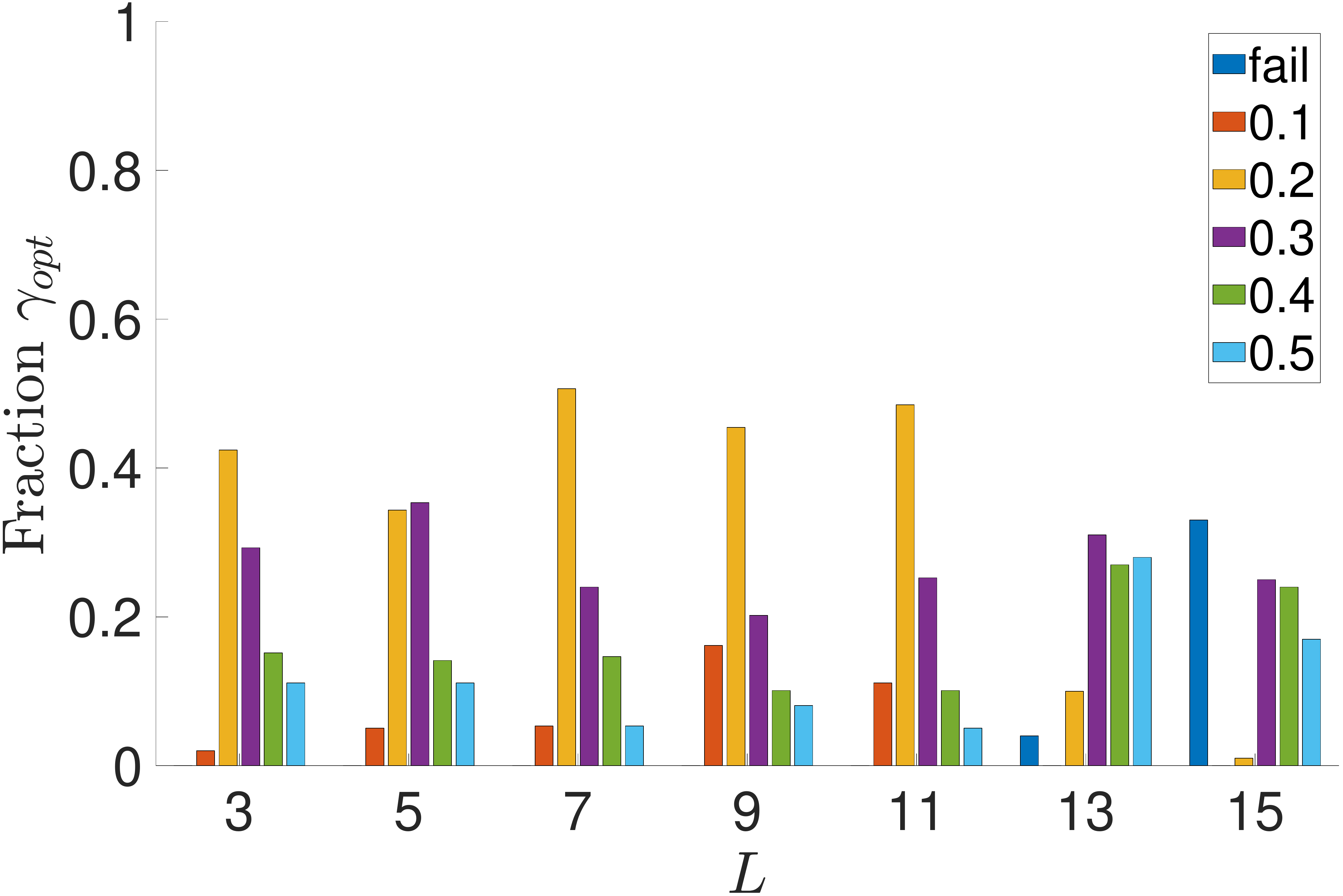}}
  \caption{\textbf{Optimal penalty strength histograms.} We show histograms of the optimal penalty strength $\gamma_{\mathrm{opt}}\in\{0.1,\dots,0.5\}$ for the set of instances at each size, including the cases when no solution was found for any strength (denoted ``fail''). As can be seen in comparing (a) to (b), the noisier problems required stronger penalties.}
  \label{fig:penalty_strengths}
\end{figure*}

\subsection{The no added noise case}
\label{sec:eta0}

The results discussed in the main text excluded the $\eta=0$ results for $L\geq 13$. We now explain why, and provide an analysis focused on the performance in the case of no added noise.

As discussed in the main text in~\cref{sec:ecn}, there is an intrinsic level of control noise. When $\eta=0$, only this noise plays a role. We show the TTS for both C and QAC in the $\eta = 0$ case in~\cref{fig:rts_noise_0}. As can be seen, there is a sudden change in C's performance when we switch from the DW2X to the DW2000Q at $L = 13$. This is consistent with there being less noise on the later generation machine, the DW2000Q. However, the latter also has a larger fraction of active qubits, which, as discussed~\cref{sec:counts}, yields a higher count of couplers involved in the problem instances for $L \geq 13$ [\cref{fig:coupling_count}]. Since the physical implementation of QAC uses four times more couplers than the C strategy, QAC should be  much more affected by noise due to this jump in coupler count. Thus, for C the lower intrinsic noise dominates, while the smooth behavior seen for QAC is likely due to a cancellation of the lower intrinsic noise with the higher noise due to the higher coupler count. This argument explains why C exhibits a discontinuity in its TTS between $L=12$ and $L=13$, and why QAC appears to transition smoothly from the DW2X to the DW2000Q.

Furthermore, this difference in physical implementation also implies that QAC will be more sensitive to coupler cross talk effects than C. Indeed, the harmful effect of cross talk can be seen in~\cref{fig:crosstalk}, in which a theoretical $3$-copies repetition code outperforms the physical implementation for every instance. Thus, the jump in coupler count from the DW2X to the DW2000Q would introduce significant cross talk that is not harming this implementation of C. The benefits of a less noisy machine are thus countered by the cross talk associated with the sharp increase in physical coupler count for QAC, while the idealized form of C simply benefits from less noise.

\section{Optimal penalty strength}
\label{app:pen}

\cref{fig:penalty_strengths} shows histograms of the optimal penalty strength defined in~\cref{eq:10}. As can be seen, as we increase the amount of noise added, the optimal penalty strength increases. This ought to be expected, since a noisier problem will require more error suppression. The DW2000Q histograms (i.e., $L\ge 13$) tend slightly more towards the larger penalty strengths once we have added enough noise. This is consistent with the discussion in~\cref{app:difference} about why the instances on the DW2000Q have more couplings for a given grid size.

\end{document}